\documentclass{article}
\usepackage{draftwatermark}
\usepackage{graphicx}
\DeclareGraphicsExtensions{.pdf,.png,.jpg}
\usepackage[symbol]{footmisc}
\usepackage{bbold}
\usepackage[fixed]{fontawesome5} 
\usepackage{float,subfig}
\usepackage{natbib}
\usepackage{comment,fullpage}
\usepackage{amsmath,amsfonts,amsthm,amssymb}
\usepackage{mathtools}

\usepackage{color}
\numberwithin{equation}{section}

\newcommand{\Vcal}{\mathcal{V}} 
\newcommand{\Hcal}{\mathcal{H}} 
\newcommand{\Scal}{\mathcal{S}} 
 
\newcommand{\Ntot}{N_\mathrm{tot}} 
\newcommand{\neigh}{\mathrm{neigh}} 
\usepackage{authblk}
\usepackage{soul}
\usepackage{cleveref} 
\usepackage{upgreek}   
\usepackage{mathastext} 
\begin{document}

\title{Multicellular Tumour Spheroids Exposure to Pulsed Electric Field: A Combined Experimental and Mathematical Modelling Study Highlighting Temporal Dynamics of DAMP Release and Accelerated Regrowth at Intermediate Field Intensities}

\author[1,$\star$,$\dagger$]{E. Leschiera}

\date{\small{$^\star$ EL and NM contributed equally to this work\\
$^\dagger$ Correspondence should be addressed to Emma Leschiera (emma.leschiera@devinci.fr), Muriel Golzio (muriel.golzio@ipbs.fr), Jelena Kolosnjaj-Tabi (jelena.kolosnjaj-tabi@ipbs.fr) or Clair Poignard (clair.poignard@inria.fr).}}

\author[2,$\star$]{N. Mattei}
\author[2]{M.-P. Rols}
\author[2,$\dagger$]{M. Golzio}
\author[2]{J. Kolosnjaj-Tabi}
\author[3,$\dagger$]{C. Poignard}

\affil[1]{De Vinci Higher Education, De Vinci Research Center, Paris, France}
\affil[2]{Institut de Pharmacologie et de Biologie Structurale, Universit\'e de Toulouse, CNRS, 	UPS, 31077 Toulouse, France}
\affil[3]{AIMOKA: Inria, Assistance Publique-Hôpitaux de Paris, Université Sorbonne Paris Nord, F-35042 Rennes, France}


\maketitle
\begin{abstract}
Electroporation is increasingly used as a percutaneous ablation technique for tumours located near vital structures. Although effective, tumour regrowth may still occur. At the same time, \textit{in vitro} studies on cell monolayers have shown that electroporation can trigger immunogenic cell death (ICD) through the release of damage-associated molecular patterns (DAMPs). These molecules can stimulate the immune system and could counteract tumour regrowth. To fully exploit electroporation, two key questions must be addressed: (1) what dynamics drive tumour regrowth, and (2) how ICD unfolds in space and time within three-dimensional cellular structures, which better mimic \textit{in vivo} conditions than 2D cultures. Here, we combine \textit{in vitro} experiments with a hybrid individual-based/continuous computational model to explore tumour spheroid regrowth and ICD potential under different pulse intensities. Experiments quantify spheroid viability, growth rate, and the release of ATP and HMGB1. In parallel, the hybrid model predicts the dynamics of proliferative, quiescent, and necrotic cells. Both approaches show that cell death and DAMP release scale with pulse intensity. The model, validated against experimental data, further highlights the dual role of quiescent cells: some die and free space and resources, while others survive and resume proliferation. Together, these findings demonstrate how spheroid fate depends on pulse strength and emphasize the importance of accounting for quiescent cells when designing electroporation-based therapies.

\end{abstract}


\section{Introduction}
Irreversible electroporation (IRE), an emerging non-thermal ablation technique, is appealing to ablate unresectable solid tumours located near vital structures. It  has demonstrated favourable safety and efficacy in clinical settings \cite{cannon2013safety,niessen2017percutaneous,zimmerman2017irreversible} even for noncancerous pancreatic tumours \cite{Tasu2023}. IRE relies on electric pulses to alter the cell transmembrane voltage and causes nanometer-sized membrane defects or pores in the cells, which leads to the loss of cell homeostasis and ultimately results in cell death \cite{al2007tumor}. Depending on the pulse parameters, such as  amplitude, duration and  pulse repetition rates, irreversible or transient defects are induced in cell membranes. In clinical settings, the tissue heterogeneity and the misplacement of electrodes, often due to anatomical limitations, can result in inefficient coverage of tumour cells by the irreversibly electroporating  electric field. As a result, treatment failure may occur, eventually leading to patient relapse \cite{edd2007mathematical, golberg2015tissue, mathy2020impact}. This crucial challenge necessitates the integration of complementary strategies alongside electroporation to mitigate potential treatment failures. Among potential approaches, leveraging the patient's immune response appears particularly promising. Therefore, to optimize the therapeutic efficacy of immunotherapies, fundamental studies are needed to elucidate the spatio-temporal dynamics of immunogenic cell death (ICD). Such knowledge will ultimately inform the rational design and scheduling of immunotherapeutic interventions.

Recent studies have shown that IRE  induces ICD after the pulse exposure. This form of cell death is known to trigger the regulated activation of the immune response~\cite{dai2021irreversible, li2012immunologic, zhang2022irreversible, zhao2019irreversible}.   
 In particular, damaged, dying or dead tumour cells communicate a state of danger to the organism by passively or actively releasing damage-associated molecular patterns (DAMPs)~\cite{polajzer2020analysis},
 {\it i.e.} endogenic danger molecules. Examples of DAMPs released from cells into extracellular space are adenosine triphosphate (ATP) and high
 mobility group box~1~(HMGB1) proteins, which  have been demonstrated to trigger innate and adaptive immune
 responses via various pathways.  
  In fact, the release of these DAMPs,  accompanied by cytokines, chemokines and other inflammatory mediators, stimulates the immune response through promoting the release of pro-inflammatory mediators and recruiting immune cells (dendritic cells, macrophages, T cells and neutrophils)~\cite{fucikova2015prognostic}. While it was previously reported that electroporation promotes the release or exposure of DAMPs (namely ATP and calreticulin) from cancerous cells~\cite{polajzer2020analysis}, the observations~\cite{polajzer2020analysis} were made in models of homogeneously suspended cells, which were in direct contact with the surrounding medium. In tissues such as tumours \cite{he2021role}, which are dense and multilayered, it was also reported that the release of DAMPs, namely HMGB1, increased significantly after IRE, yet within these \textit{in vivo} experiments \cite{he2021role}, the spatio-temporal profiles of HMGB1 release was not correlated with the time after treatment. Although DAMP secretion is recognized as a hallmark response to IRE, additional studies are required to elucidate its spatio-temporal patterns in more physiologically relevant, structured environments.  
In order to increase the complexity of the \textit{in vitro} model, we thus used 3D cancer cell-derived spheroids. Spheroids are three-dimensional structures composed of cancer cells that can accurately reproduce the behaviour of small solid tumours in their preliminary avascular stage. 
They closely resemble solid micro-tumours in many aspects, such as internal gradients of pressure, nutrients, and oxygenation which induce the establishment of a proliferation gradient with an increasing proportion of quiescent and necrotic cells across microtissue depth. Precisely, in small spheroids (generally with diameters smaller than 300 \textmu m), the diffusion of nutrients up to the center of the spheroids enable them to proliferate leading to an exponential phase of the growth. As a spheroid increases in size, the cell density at its centre increases and may eventually become too high to support further proliferation, and the nutrient diffusion is hampered,  driving cells to stop division and enter into a quiescent state. Slower growth of the spheroids will occur until nutrient levels at its centre fall below those needed to maintain cell viability, leading to the formation of a central necrotic core composed of dead cells.  These similarities with small solid tumours provide great potential for studying the intrinsic properties of tumours, thus they are often used for drug screening, as well as to determine the therapeutic efficacy. Such models demonstrated their relevance to address the electroporation phenomena \cite{frandsen2015calcium, gibot2013antitumor,wasungu20093d}.

In parallel, there is an increasing interest in using mathematical modelling to describe the growth of tumour spheroids and their response to therapy.  The Logistic and Gompertz models, which are two classical ordinary differential equation (ODE) models for tumour growth \cite{benzekry2014classical,norton1988gompertzian, spratt1993decelerating,Vaghi2020}, recapitulate the characteristic sigmoid curve describing how the total spheroid volume changes over time. Usually, these models represent tumour spheroids as homogeneous entities, which limit the description of their internal spatial structure. More detailed mechanistic models based on partial differential equations (PDEs) relate the internal spatial structure of the spheroids to the supply of vital elements such as oxygen and nutrients \cite{collin2022spatial, greenspan1972models, michel2018mathematical}. These more complex approaches involve additional parameters but enable representing the heterogeneity of the response of the tumour cells within the tumour spheroids. However, differential equation models are defined on
the basis of population-level phenomenological assumptions,
which may limit the level of biological detail that can be included
in the model.
By using computational models, such as stochastic 
individual-based models (IBMs), a more direct and precise mathematical
representation of tumour spheroid dynamics can be achieved  \cite{bull2020mathematical, caraguel2016towards, leschiera2023individual}. These models track the dynamics of single cells, thus permitting the representation of single cell-scale
mechanisms, and account for possible stochastic fluctuations in single-cell biophysical
properties. IBMs are often multiscale, linking processes that act at the tissue, cell and subcellular scales. 
IBMs are termed ‘hybrid’ if they combine different modelling approaches. For example, a PDE describing the spatial distribution of nutrients within a tumour spheroid may be coupled to a stochastic IBM describing the dynamics of individual tumour cells.

Our objective here is twofold. From the biological view-point, we applied different intensities of electric field to the spheroid model, in order to mimic the heterogeneity of electric fields in clinical situations. Spheroid growth rate and viability, as well as dynamics of ATP and HMGB1 release were assessed. Importantly, our data indicate a delayed cell death and HMGB1 release followed by a strong regrowth in the 1500 V/cm IRE condition, resulting in partial ablation. From the mathematical modelling view-point, we used the biological results, taking advantage of the spatial dimension of our 3D model, as the basis to develop a spatial IBM model describing spheroid dynamics after electroporation. First, the model describes the growth of spheroids, regulated by a nutrient supply, prior to electroporation. Based on the gradients of proliferative, quiescent, and necrotic cells, this model reproduces the experimentally observed growth patterns of untreated tumour spheroids. Then, using the electric field intensities considered in the experiments, our hybrid IBM studies the impact of the different pulse intensities on cell death, spheroid regrowth, and DAMP release. The numerical results are also qualitatively validated, namely, confronted to additional \textit{in vitro} biological data. Our IBM provides new insights into the spatio-temporal evolution of cell death and DAMP release. In particular, in the IBM, the dynamics of spheroid regrowth is explained by the fate of its quiescent cells. The ones dying from the pulse exposure will free up space and resources, thereby promoting the regrowth of surviving proliferative cells, which resume the cell cycle, ultimately accelerating spheroid regrowth.

The subsequent sections are organised as follows. Materials and methods are described in Section \ref{Material and method}, where we focus on \textit{in vitro} experiments that motivated our modelling assumptions, and Section \ref{Modelling framework}, where we elaborate the hybrid IBM. Details of model parameterisation and sensitivity analysis are provided in Appendix \ref{calibration} and Appendix \ref{sensitivity analysis}, respectively. 
 Section \ref{in vitro_Results} reports the main \textit{in vitro} experimental results, while the results of numerical simulations and their experimental validation are presented in Section \ref{Numerical_results}. Section~\ref{discussion} discusses numerical results in the context of the experimental observations and presents some research perspectives. Section \ref{conclusion} concludes the paper.


 \section{Materials and Methods}
\subsection{\textit{in vitro} experiments}
\label{Material and method}

\paragraph{Cell Culture} Biological experiments involved the murine hepatocarcinoma cell line Hepa 1-6 (ATCC\textsuperscript{\textregistered} CRL-1830\texttrademark) and the Hepa 1-6 stably expressing the green fluorescent proteins (GFP), which was used for spheroid growth assay. Cells were grown under standard culture conditions (5\% CO$_2$, 37$^\circ$C) in the Dulbecco’s Modified Eagle Medium (DMEM, Gibco-Invitrogen, Carlsbad, CA, USA) containing 4.5 g/L glucose, L-Glutamine and pyruvate, 100 U/mL penicillin, 100 \textmu g/mL streptomycin, and 10\% of fetal bovine serum. Cancer cell spheroids  were produced by seeding 3000 cells in 200 \textmu L of culture medium per well in round bottom Costar Corning Ultra-low attachment 96-well plates (\#7007, Fisher Scientific, Illkirch, France). Plates were maintained in 5\% CO$_2$ humidified atmosphere at 37$^\circ$C. After four days of culture, spheroids were considered suitable for treatment.

\paragraph{Electroporation}
Electroporation was achieved following the delivery of 80 unipolar pulses at 0 V/cm (control), 500 V/cm, 1000 V/cm, 1500 V/cm, 2000 V/cm or 2500 V/cm. The mathematical model was calibrated using the 0, 500, 1500 and 2500 V/cm conditions, while the results with 1000 and 2000 V/cm were used to validate the model. The duration of a pulse was 100 \textmu s, and the pulses were applied at a pulse repetition rate of 1000 Hz. The pulses were generated by the Electrocell S20 generator (Leroy Biotech, St Orens, France). The pulses were applied to multicellular spheroids in a low conductivity iso-osmotic pulsing buffer  (8.1 mM dipotassium phosphate, 1.9 mM monopotassium phosphate, 1 mM magnesium chloride and 250 mM sucrose in water; pH: 7.4, osmolarity: 270 osmol/L, conductivity: 1.7 mS/cm). Three spheroids were treated simultaneously in a drop of 100 \textmu L buffer, placed on a Petri dish between parallel plate stainless steel electrodes with an inter-electrode distance of 0.4 cm (Supporting Figure \ref{supplementary2}).  
The values of the electric field intensity (in V/cm) correspond to the voltage between the electrodes divided by their distance.  The temperature increase due to electroporation was measured with the fiber-optic temperature sensor (Nomad Fiber Optic Thermometer Model NMD 535 A). In this setting, the maximum temperature increase was of~$\sim17^\circ$C, (at the starting temperature of 25$^\circ$C a maximum of 42$^\circ$C was attained), with a fast decrease over the next seconds (8$^\circ$C decrease in less than 10 seconds) (Supporting Figure \ref{supplementary3}), reaching the basal temperature after 70 to 80 seconds. Spheroids were treated four days after seeding (considered as Day 0), and subsequently, 10 minutes after treatment, replaced in culture medium for longitudinal follow-up and further analysis.

\paragraph{Viability of the spheroids}
Spheroids were monitored by fluorescence and bright field videomicroscopy at the interval of 24h for 4 days following pulse treatment, using the IncuCyte Zoom Live Cell Analysis System Microscope (Essen BioScience IncuCyt, Herts, Welwyn Garden City, UK) at 10x magnification.For GFP-expressing spheroids, green fluorescence served as a marker of viability. To assess cell death, 1 \textmu M propidium iodide (PI) was added to the culture medium of treated spheroids. The GFP-positive area was quantified to determine the size of the viable spheroid region. 
	
	Apoptosis was evaluated by caspase 3/7 activity using the NucView\textsuperscript{\textregistered} 488 Caspase-3 Assay Kit for Live Cells (Biotium, Froment, USA). Briefly, non-GFP Hepa 1-6 spheroids were incubated with 1 \textmu M of the caspase-sensitive fluorescent probe for 24 hours after treatment. GFP and NucView signals were detected with a green filter (Excitation 460/20 nm, Emission 524/20 nm), while PI staining was detected with a red filter (Excitation 565/605 nm, Emission 665/40 nm). 


\paragraph{Image analysis}
Analyses of micrographs were performed on ImageJ software (\textquotedblleft National Institutes of Health\textquotedblright (USA) software version 1.54f). GFP-positive areas were measured using a threshold method: first, a threshold of minimum GFP intensity was applied on each image, then areas above this threshold were measured. Growth speed of each spheroid was computed as the slope of the linear regression line between the areas measured at Day 1 and Day 4. Additionally, the growth speed was measured before spheroid treatment between Day -2 and Day -1. Caspase 3/7 activation was determined by measuring the mean intensity of fluorescence intensity of NucView 488\textsuperscript{\textregistered} within each spheroid.

\paragraph{ATP release}
ATP released from a pool of 3 spheroids was assayed 10 minutes and 24 hours after treatment using CellTiter-Glo 2.0 Luminescent Cell Viability Assay (Promega, Madison, WI, USA). Briefly, the pulse buffer was collected and incubated with the kit reagent for 10 minutes in the dark at room temperature. Then, luminescence was measured using a plate reader (CLARIOstar, BMG Labtech, Oertenberg, Germany). Additionally, Intracellular ATP of untreated spheroids was assayed by lysing spheroids with the kit reagent. Luminescence measurements were reported relative to the 0 V/cm condition.

\paragraph{HMGB1 release}
HMGB1 released  from a pool of 27 spheroids was assayed 3 and 6 hours after treatment by immunoblotting using an antibody against HMGB1 (3935, Cell Signalling Technology, Danvers, USA). Briefly, 3 or 6 hours after treatment, the culture media of spheroids were collected. Proteins were precipitated with glacial acetone and incubated in buffer with a protease inhibitor cocktail (BS386, Bio Basic Inc., New York, USA). Total protein concentration was measured by Bradford assay, and the same level of proteins was run on 4-20\% Mini-PROTEAN TGX Stain-Free gels (Bio-Rad, Hercules, USA) before being transferred to a nitrocellulose membrane. The proteins were probed with the HMGB1 antibody according to the manufacturer’s instructions, followed by incubation with an anti-rabbit HRP-conjugated secondary antibody (A120-111P, Axil Scientific, Singapore). Protein revelation was performed using the Western blot imaging ChemiDoc\texttrademark\ Touch Imaging System (Bio-Rad, Hercules, USA) with enhanced chemiluminescence. Total protein obtained by stain-free imaging was used to normalize each band, and analyses were performed using Image Lab\texttrademark\ Software (version 6.0.1, Bio-Rad, Hercules, USA). Each band intensity was reported relative to the 2500 V/cm condition.

\paragraph{Data analysis}
Biological data were generated from three independent experiments (N = 3), each comprising a minimum of six technical replicates (6 multicellular spheroids); or as specified in the respective figure captions in dedicated Results sections. Statistical analyses were performed using GraphPad Prism version 5.04 (GraphPad Software, Inc., La Jolla, CA, USA), and data in the graphs are expressed as mean ± standard error of the mean (SEM). One-way ANOVA tests followed by a Dunnett’s multiple comparison test or two-way ANOVA tests followed by a Bonferroni multiple comparison test were performed. Overall statistical significance was set at $p< 0.05$.

\paragraph{pH measurement}
The change in the pH of the pulsation buffer (low conductivity iso-osmotic pulsing buffer constituted of 8.1 mM dipotassium phosphate, 1.9 mM monopotassium phosphate, 1 mM magnesium chloride and 250 mM sucrose in water; pH: 7.4, osmolarity: 270 osmol/L, conductivity: 1.7 mS/cm) was measured with a pH probe (InLab Micro, Mettler Toledo, USA). Precisely, 150 µl of pulsing buffer (ZAP) or 100 µL of pulsing buffer with 50 µl of water (ZAP + H$_2$O) were pulsed under the same condition as the spheroids. Three measurements were conducted per condition.

\subsection{Mathematical model}
\label{Modelling framework}
In this section, we present the mathematical model. In the first part, we introduce the hybrid IBM used  to model the growth of the spheroids in normal conditions when electroporation is not applied (control group). We then extend the model to include the electric field considered in the experimental set-up, and study its effects on spheroid loss of viability, regrowth, and the release of ATP and HMGB1.

The computation domain partitioned by a Cartesian grid. Each node of the grid is occupied by a number of cells, which are viewed as agents following  biological laws. We discretise the time variable $t$ as $t_k =k \tau$ with $k \in \mathbb{N}$ where $\tau \in \mathbb{R^*_+}$  is the time step. In the domain $\Omega=(-\ell/2,\ell/2)^d$, with $d=2$ or 3, depending on the computational setting. The space variable $\mathbf{x_i}\in\Omega$ is defined on the lattice as follows, for any $d$-uple $\mathbf{i}\in [0, \mathcal{N}]^d \subset \mathbb{N}^d$, $\mathbf{x_i}=\chi\mathbf{i} -\ell/2$ where $\chi \in \mathbb{R^*_+}$ is the path of discretisation of the Cartesian lattice. Note that $\mathcal{N} := 1 + \lceil\frac{\ell}{\chi}\rceil$ , where
$\lceil\cdot \rceil$ denotes the ceiling function. Here, $\mathbb{N}$ is the set of natural numbers including zero. 

The tumour spheroid is composed of three groups of cells (also known as cell state transitions): proliferative, quiescent, and necrotic cells. Proliferative cells divide at a rate dependent on the local nutrient concentration, described by a discrete, non-negative function. When a tumour cell divides, it produces two daughter cells; one remains in the parent cell's position, while the other occupies either the same position or an adjacent empty space. If proliferative cells lack the space needed to divide, they cease proliferating and enter the quiescent state. Finally, when nutrient concentration is too low, both proliferative and quiescent cells become necrotic. The impact of the electric field on the spheroid is twofold. On the one hand it generates immediate cell death, and on the other hand it lead to release of ATP and HMGB1, whose amount is described as a function of the electric field intensity.

\subsubsection{Spheroid growth modeling prior electroporation}
\label{growthPriorElect}
We denote by $n_\mathbf{i}^k$ the density of tumour cells, which is defined as the number of tumour cells at position $\mathbf{x_i}$ and at time $t_k$, $N_\mathbf{i}^k \in \mathbb{N}$, divided by the size of a voxel (pixel if $d=2$), that is
\begin{equation}
	n_\mathbf{i}^k\equiv n(\mathbf{x_i},t_k):=\frac{N_\mathbf{i}^k}{\chi^d}.
	\label{eq:tumourDensity}
\end{equation}
We respectively denote by $p_\mathbf{i}^k, q_\mathbf{i}^k$ and $r_\mathbf{i}^k$ the density of proliferative, quiescent and necrotic cells and we assume that
\begin{equation}
p_\mathbf{i}^k +q_\mathbf{i}^k+r_\mathbf{i}^k= n_\mathbf{i}^k.
\end{equation} 
The total numbers of tumour cells at time $t_k$ is denoted by $\Ntot^k=\sum_{\mathbf{i}}N^k_{\mathbf{i}}$. We denote by $\Vcal^k$ the spheroid volume at time $t_k$, which is defined as 
\begin{equation}
\Vcal^k=\chi^d\sum_{\mathbf{i}}\Hcal(N_\mathbf{i}^k-N_\Scal),
\label{AreaSpheroide}
\end{equation} 
where $\Hcal$ is the usual Heaviside function , $\Hcal(x)=\begin{cases}0,\text{if $x\leq0$}\\1,\text{if $x>0$}\end{cases}$. The parameter $N_\Scal\geq 0$ enables us to exclude positions where the density of tumour cells is too low.
\paragraph{Nutrient concentration}
 Let $C$ be the nutrient concentration. We assume that nutrients are  consumed by proliferative cells $p$ at rate $\alpha_{C}\in\mathbb{R}^*_+$ and by  quiescent cells $q$ at a lower rate $\eta_C:=c_1\alpha_{C}\in\mathbb{R}^*_+$, where $c_1\in\, ]0,1[$. The equation governing the spatio-temporal evolution of the nutrient concentration at position $x_\mathbf{i}$ and at time $t_k$ is thus
\begin{equation}
\begin{cases}	
&\partial_tC -\beta_C\nabla^2C=-\alpha_{C}(p+c_1q)C, \quad \text{in $\Omega$},\\
&C|_{t=0}=C_{\rm out},\quad C|_{\partial\Omega}=C_{\rm out},
\end{cases}
	\label{nutrent}
\end{equation}
where $\beta_C\in\mathbb{R}^*_+$ is the nutrient diffusion rate and $C_{\rm out}$ is the uniform concentration at the initial time of the experiment.
The above equation is discretised with the standard second order approximation in space, and explicit forward Euler scheme in time:
\begin{align}
&C^0_{\mathbf{i}}=C_{\rm out},
\\
&C^{k+1}_\mathbf{i}=(1-\tau \alpha_{C}(p^{k}_\mathbf{i}+c_1q^{k}_\mathbf{i}))C^{k}_\mathbf{i} +\frac{\tau \beta_C}{\chi^2} \sum_{\mathbf{j}\in\neigh(\mathbf{i})}(C^{k}_\mathbf{j}-C^{k}_\mathbf{i}),
\end{align}
where $\neigh(\mathbf{i})$ is the set of the $2d$  closest neighbouring indices of $\mathbf{i}$. Note that the above discretization leads to a Courant-Friedrichs-Lax condition $\frac{\tau \beta_C}{\chi^2}<1$ which in our case is not restrictive, but static version of equation \eqref{nutrent} could also be used without affecting the results. 

\paragraph{Tumour cell proliferation.}
\label{Tumour cell proliferation and natural death}
At the initial time of simulations, (\textit{i.e.} at day -1) a predetermined number of tumour cells are positioned in the centre of the domain. Initially, all tumour cells are in the proliferative state $(p^0_{\mathbf{i}},q^0_{\mathbf{i}},r^0_{\mathbf{i}})=(n^0_{\mathbf{i}},0,0)$, $n^0_{\mathbf{i}}$ being given at any grid point on the lattice. At every time-step $k\geq1$, a proliferative cell on grid-site $\mathbf{i}$  will undergo cell division with probability
\begin{equation}
	\tau\alpha_p \frac{C_\mathbf{i}^k}{C_{\text{max}}} >0,  
	\label{eq:tumProl}
\end{equation}
where $\alpha_p\in\mathbb{R^*_+}$ 
represents the rate of tumour cell proliferation, $C_\mathbf{i}^k$ is the concentration of nutrient at position $x_\mathbf{i}$ and at time $t_k$ and $C_{\text{max}}\geq C_\mathbf{i}^k$ is the needed nutrient concentration for proliferation. 

The growth of proliferative cells is modelled by assuming that cell proliferation may induce a passive movement of daughter cells. Therefore,  one of the daughter cells  will occupy  the  same  position  as  the  parent  cell,  while  the  other  will be randomly placed either at the same site of the parent cells or in one of the $2d$ neighbouring grid points. 
\paragraph{Tumour cell quiescence due to tight cell packing.} 
It is known that tight cell packing induce a switch from a proliferating to a quiescent state of cancer cell.  To model this behaviour, we restrict tumour cell proliferation to spatial locations $\mathbf{x_i}$ where the tumour cell density $n_\mathbf{i}^k$ is below a specified threshold value $n_{\max} \in \mathbb{R}^*_+$, representing a state of tight cell packing. When $n_\mathbf{i}^k = n_{\max}$, and the density of tumour cells in the lattice neighbouring sites also reaches the threshold value $n_{\max}$, the proliferative cells at position $\mathbf{x_i}$ stop proliferation and enter a quiescent state. 
Note that if the tumour cell density $n_\mathbf{i}^k$ once again falls below $n_{\max}$ (for instance, following electroporation), quiescent cells resume their proliferation state.

\paragraph{Tumour cell necrosis.}

Under hypoxia condition, which often coincide with local nutrient depletion, tumour cells enter the necrotic state. We denote by $\gamma$ the function which describes the switch rate at which proliferative and quiescent cells become necrotic. Following \cite{michel2018mathematical}, we define $\gamma(C)$ by using the following formula:
\begin{equation}
	\gamma(C)=\gamma_{\min}+\frac{\gamma_{\max}-\gamma_{\min}}{2}(1-\text{tanh}(K(C-C_{\text{hyp}})), 
	\label{switching}
\end{equation}
where $ \gamma_{\min}$ and  $ \gamma_{\max}$ stand for the infemum and the supremum of $\gamma$ (these values are not reached), $C_{\text{hyp}}$ is a hypoxia threshold, and $K$ is a parameter which controls the stiffness of $\gamma(C)$.
\subsubsection{Release of ATP and HMGB1 after electroporation}
In this section, we extend the hybrid IBM to take into account the death of the spheroids due to electroporation,  their regrowth and their release of ATP and HMGB1. Note that here, to simplify our model, we do not explicitly model electroporation dynamics; instead, we directly use the electric field intensities considered in the experimental set-up.

 \paragraph{Release of ATP.}
\label{Dynamic of for the chemoattractant}
After the pulse delivery, tumour cells rapidly release ATP into the supernatant, and the ATP release increases with higher initial pulse intensity (see for example  \cite{polajzer2020analysis}).
In the experiments presented in Section \ref{Material and method}, the measurements focus on the amount of ATP in the supernatant 10 minutes after the pulse.  Therefore, to reduce
the number of parameters and the complexity of the model, we do not explicitly model the dynamics of ATP release. Instead,  we assume that the total amount of ATP released increases with the intensity of the electric field.

Let $S$ denote the total amount of ATP released in the supernatant and $\bold{E}$ the electric field intensity. We define $S$ as a polynomial function of the electric field $\bold{E}$ with a maximum degree of three, i.e. 
\begin{equation}
	 S(|\bold{E}|) = \beta_0 + \beta_1 |\bold{E}| + \beta_2 |\bold{E}|^2  +\beta_3 |\bold{E}|^3,
\label{ATP}
\end{equation}
where the effective degree of the polynomial, as well as the coefficients $(\beta_0,\ldots, \beta_3) \in\mathbb{R}^4$ will be determined in accordance with the experimental data using suitable statistical tests. 

\paragraph{Loss of viability due to ATP release.}
 It has been shown that as tumour cells undergo various stress conditions, including therapeutic interventions such as electroporation, the amount of ATP released plays a significant role in triggering apoptotic pathways \cite{jakstys2020correlation,rols1990electropermeabilization}. Therefore, in our model we assume that tumour cell death due to electroporation is proportional to the total amount of ATP released in the supernatant. 
In the literature, studies have demonstrated that the duration of cell death can vary depending on the intensity of electric pulses \cite{napotnik2021cell}. Specifically, higher intensities of electric pulses may induce a more rapid cell death response, while lower intensities may result in a slower or delayed cell death process.  Accordingly, we require that the duration before cell death takes place is inversely proportional to the intensity of the electric field applied, \textit{i.e.} the higher the electric fields, the faster  the cell death. 
 
 Let $Z$ be a random variable drawn from a uniform distribution with bound parameters $S\pm 10\%S$ (\textit{i.e.} $Z\sim \mathcal{U}_{[S-10\%S,S+10\%S]}$). When electroporation is applied, the magnitude of the electric field $\bold{E}$ predetermines the cells which will die after the pulse. The  released ATP determines the speed of death. More precisely,  at every time-step $k$, proliferative or  quiescent cells, which have been subjected to a high enough electric field, will die with the following probability \begin{equation}
	\tau c_2Z >0, 
	\label{eq:tumDeath}
\end{equation} where  $c_2 \in (0,1/\tau)$ is a scaling factor. The density of the above  tumour cells, which  died by electroporation is denoted  $m_\mathbf{i}^k=m(\mathbf{x_i},t_k)$. At the end of the time-step, dead cells are then removed locally at the spatial locations where cell death occurs within the computational grid.

 In our model, we assume a fixed time window for tumour cell death following electroporation, specifically set at 6 hours. This implies that the impact of electroporation on tumour cells ceases after this predefined time frame.
\paragraph{Release of HMGB1.}
In the experiments presented in Section \ref{Material and method}, HMGB1 release is measured 3 hours and 6 hours after electroporation.

 In the literature, it has been shown that HMGB1 can be passively released after various types of cell death, including apoptosis \cite{bell2006extracellular}.   Therefore, in our model we assumed that, before they are removed from the domain, dead tumour cells secrete HMGB1. We denote by $P^k$  the total amount of  HMGB1  at time $t_k$ and by

\begin{equation}
\rho_m^k\equiv\rho_m(t_k):=\sum_im_\mathbf{i}^k\chi^d
\end{equation}
 the total number of dead tumour cells at time $t_k$.   At each time-step $t_k$, we assume that the release of HMGB1  is proportional to the total number  of dead tumour cells  $\rho_m^k$, and we model its dynamics by the following discrete balance equation
\begin{equation}
	P^{k+1}\equiv P(t_{k+1}):=P^k+\tau\alpha_{P}\rho_m^k, \quad k\in \mathbb{N}.
	\label{HMGB1}
\end{equation}
  In Eq.~\eqref{HMGB1}, $\alpha_{P}\in \mathbb{R}^+$ models the release of HMGB1 by dead tumour cells.

\section{Results}
\label{Results}
This section presents the main experimental and numerical results of our study. Section \ref{in vitro_Results} reports the \textit{in vitro} experimental results, while the results of numerical simulations are presented in Section \ref{Numerical_results}.
\subsection{Effects on spheroids as determined in biological experiments}
\label{in vitro_Results}
\subsubsection{Effects of electric field intensity on spheroid viability and growth}
\begin{figure}[H]
	\centering
	\includegraphics[width=14cm]{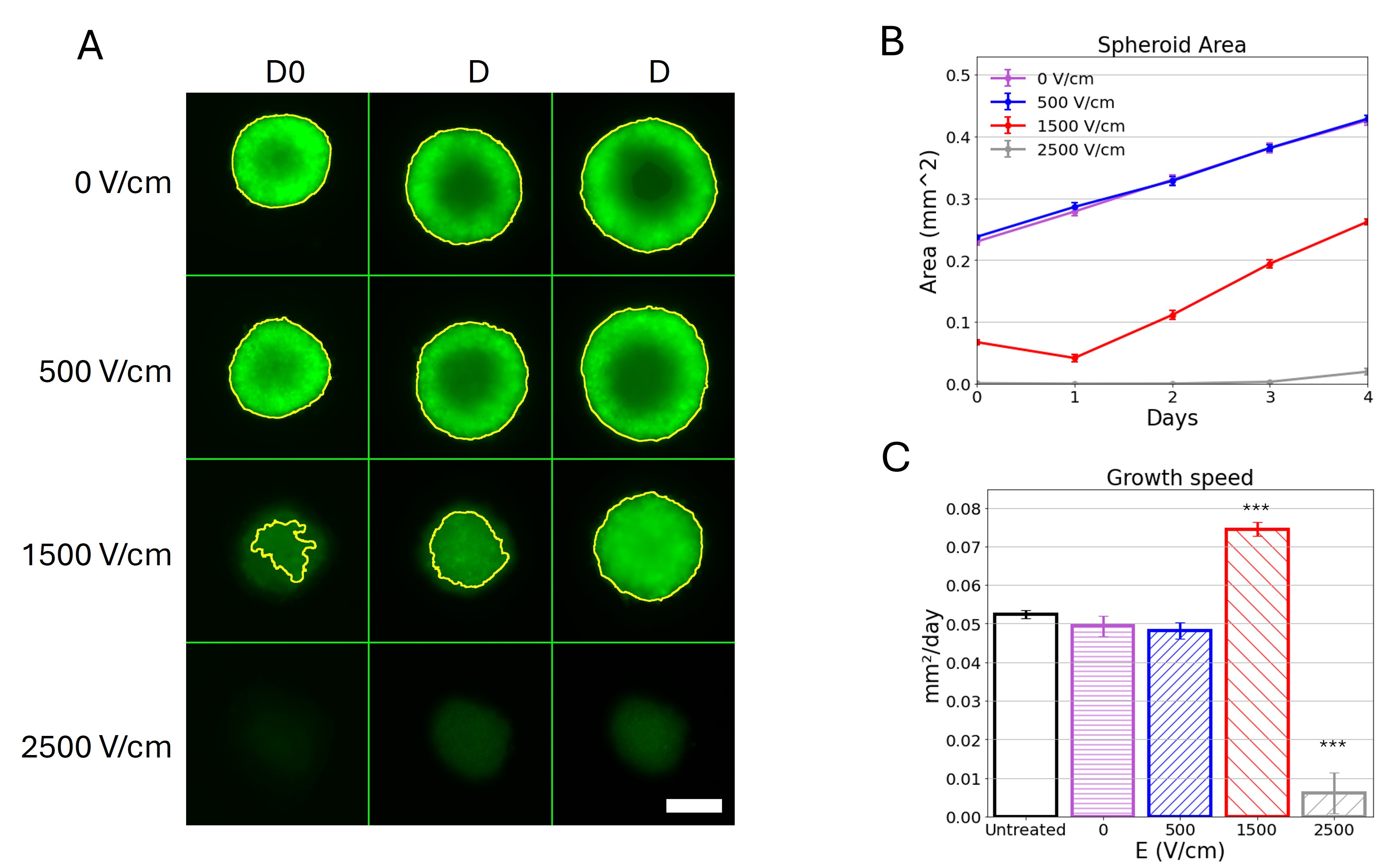}
	\caption{\textbf{Analysis of GFP-expressing Hepa 1-6 spheroids growth after electroporation treatment } (A) Representative fluorescence micrographs of spheroids at Day 0 (D0), Day 2 (D2) and Day 4 (D4) after treatment at varying electric field intensities. GFP channel is displayed in green. The yellow outlines correspond to viable (GFP) zone boundaries used for further analysis. Scale bar is 300 \textmu m. (B) Spheroid viability over time based on GFP-positive areas. The values obtained at Day 0 correspond to the first acquisition after treatment. (C) Spheroid growth speed based on viability curves from Day –2 to Day 1 (white bar) or after treatment with from Day 0 onwards, as per applied protocol (0, 500, 1500 or 2500 V/cm). Statistical significance was assessed using one-way ANOVA followed by a Dunnett’s multiple comparison test, in comparison to the 0~V/cm condition. *** $p < 0.001$. Data are shown as the mean ± SEM from three independent experiments (N=3), each including at least 6 spheroids per condition ($n\geq 6$).
	} 
	\label{damps_release}
\end{figure}

\begin{figure}[H]
	\centering
	\includegraphics[width=14cm]{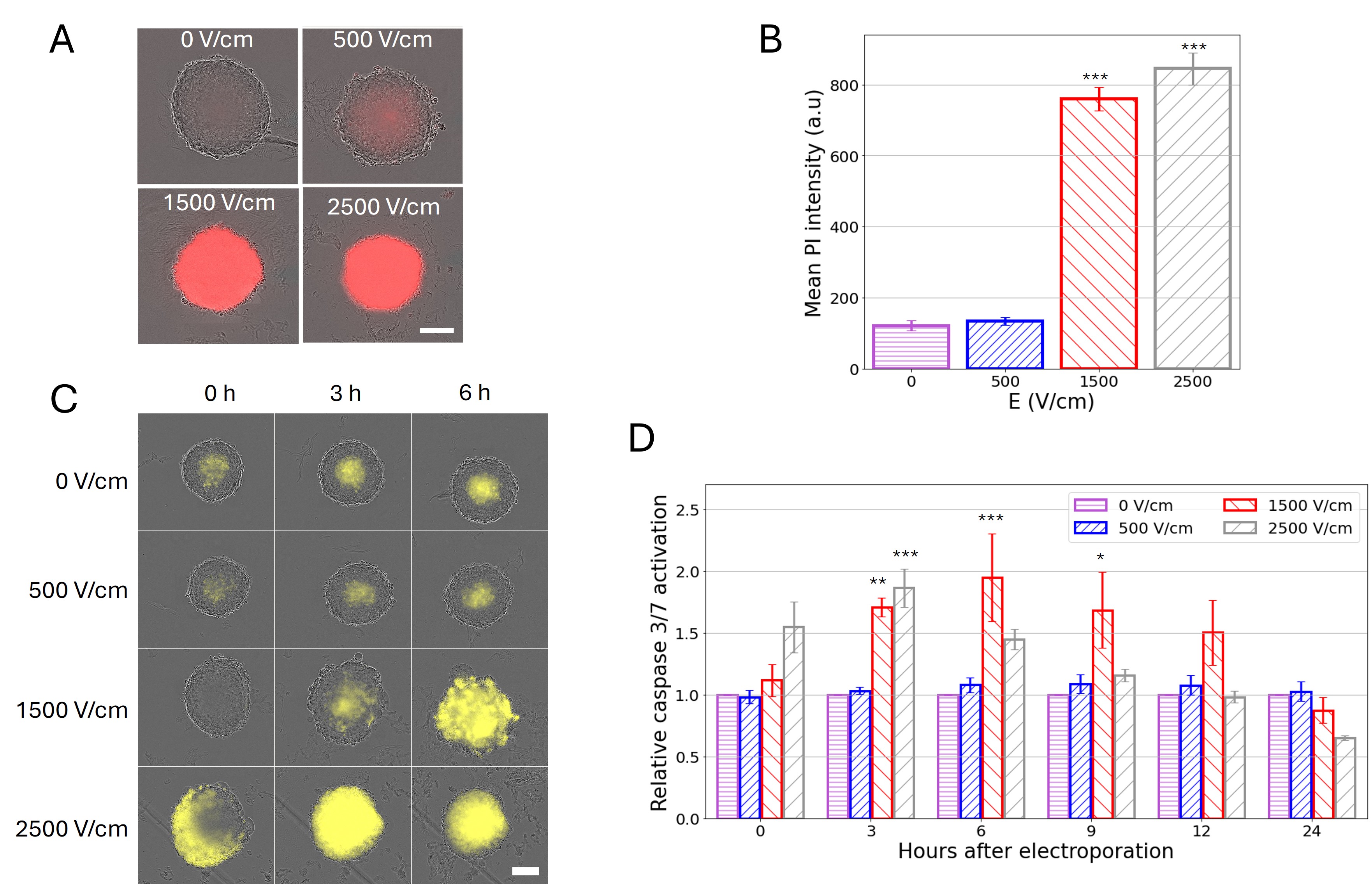}
	\caption{\textbf{Spheroid cell death after pulse treatment.} (A) General assessment of cell death irrespective of the death pathway: representative bright field and fluorescence micrographs of spheroids one day after treatment. Dead cells internalized propidium iodide and are represented in red. Scale bar is  200 \textmu m. (B) Cell death in spheroids one day after treatment, based on mean propidium iodide staining inside the spheroids. Statistical significance was assessed using an one-way ANOVA followed by a Bonferroni’s multiple comparison test, in comparison to the 0 V/cm condition. *** $p < 0.001$. (C) Representative brightfield and fluorescence micrographs of spheroids 0, 3 and 6 hours after treatment. Caspase 3/7 staining is in yellow. Scale bar is 300 \textmu m. (D) Apoptosis induction based on caspase 3/7 activation, relative to the 0 V/cm condition. Statistical significance was assessed using a two-way ANOVA followed by a Bonferroni multiple comparison test, in comparison to the 0 V/cm condition. * $p < 0.05$, **$ p < 0.001$, **** $p < 0.0001$. Data are shown as the means $\pm$ SEM from three independent experiments, each including 6 spheroids per condition.	
	} 
	\label{sph_death}
\end{figure}

Viable cell populations tracking and the follow up of the dynamics of spheroid growth were done by videomicroscopy over 4 days post-treatment  (Figure~\ref{damps_release}A). Pulse intensity of 500 V/cm had no impact on spheroid growth. Higher intensity electric fields (1500 V/cm) induced a clear loss of GFP immediately after pulsation, at Day 0. At later time-points (D2, D4), regrowth of the spheroids was observed in the 1500 V/cm condition, evidencing an incomplete spheroid ablation (Figure~\ref{damps_release}A). The highest electric field intensity (2500 V/cm) induced a major persistent loss of viability over a period of 4 days (Figure~\ref{damps_release}A). The GFP-positive areas of the spheroids (Figure~\ref{damps_release}A) were used to plot spheroid viability over time (Figure~\ref{damps_release}B) and to determine the speed of growth (Figure~\ref{damps_release}C). Low pulse intensity (500 V/cm) had no effect on spheroid viability and growth speed, with no difference in comparison to control spheroids (Figure~\ref{damps_release}B and C). Treatments of intermediate intensities (1500 V/cm) induced a transient decrease in viability (Figure~\ref{damps_release}B) followed by a significant increase in spheroid growth speed post-treatment (Figure~\ref{damps_release}C). This suggests a stimulatory effect on spheroid proliferation induced by these pulsing conditions. It is worth noting that this increased growth speed is higher than that of spheroids before day 4 of culture (prior to treatment) (Figure~\ref{damps_release}C). Hence, increased growth speed cannot be attributed exclusively to a smaller spheroid size. Finally, treatments at the highest intensity (2500 V/cm) dramatically reduced viability and inhibited spheroid growth post treatment.  
In addition to monitoring viability, cell death was assessed using propidium iodide (PI), added one day after treatment as a death pathway-independent marker of loss of membrane integrity (Figure~\ref{sph_death}A). The mean PI intensity was measured within the spheroids (Figure~\ref{sph_death}B). As expected, intense PI staining is observed in both the 1500 and 2500 V/cm conditions.

Subsequently we evaluated the fraction of apoptotic cells. Apoptosis, a form of regulated cell death, is mediated by a cascade of caspase activation, including caspases 3 and 7. To assess apoptosis induction after treatment, we measured the activation state of these caspases (Figure~\ref{sph_death}C). Fluorescence microscopy revealed apoptotic signals exclusively within the cores of spheroids treated with 0 and 500 V/cm conditions. At higher intensities, the caspase 3/7 staining is more intense and spread across entire spheroids. The mean intensity of the staining was measured over time (Figure~\ref{sph_death}D). In comparison with untreated spheroids, no substantial caspase activation was observed at 500 V/cm, consistent with the minimal impact of 500 V/cm pulses on spheroid viability. At 1500 V/cm, an increase in caspase activation was observed, with a peak reached 6 hours after treatment. At 2500 V/cm, we observed a rapid and intense increase in caspase 3/7 activation, with a peak attained about 3 hours after treatment. This disparity in the timing of caspase activation peaks indicates that 2500 V/cm pulses induce a rapid and more extensive loss of viability compared to milder pulsing conditions.

\subsubsection{Effects of electric field intensity on DAMP release}
\begin{figure}[H]
	\centering
	\includegraphics[width=14cm]{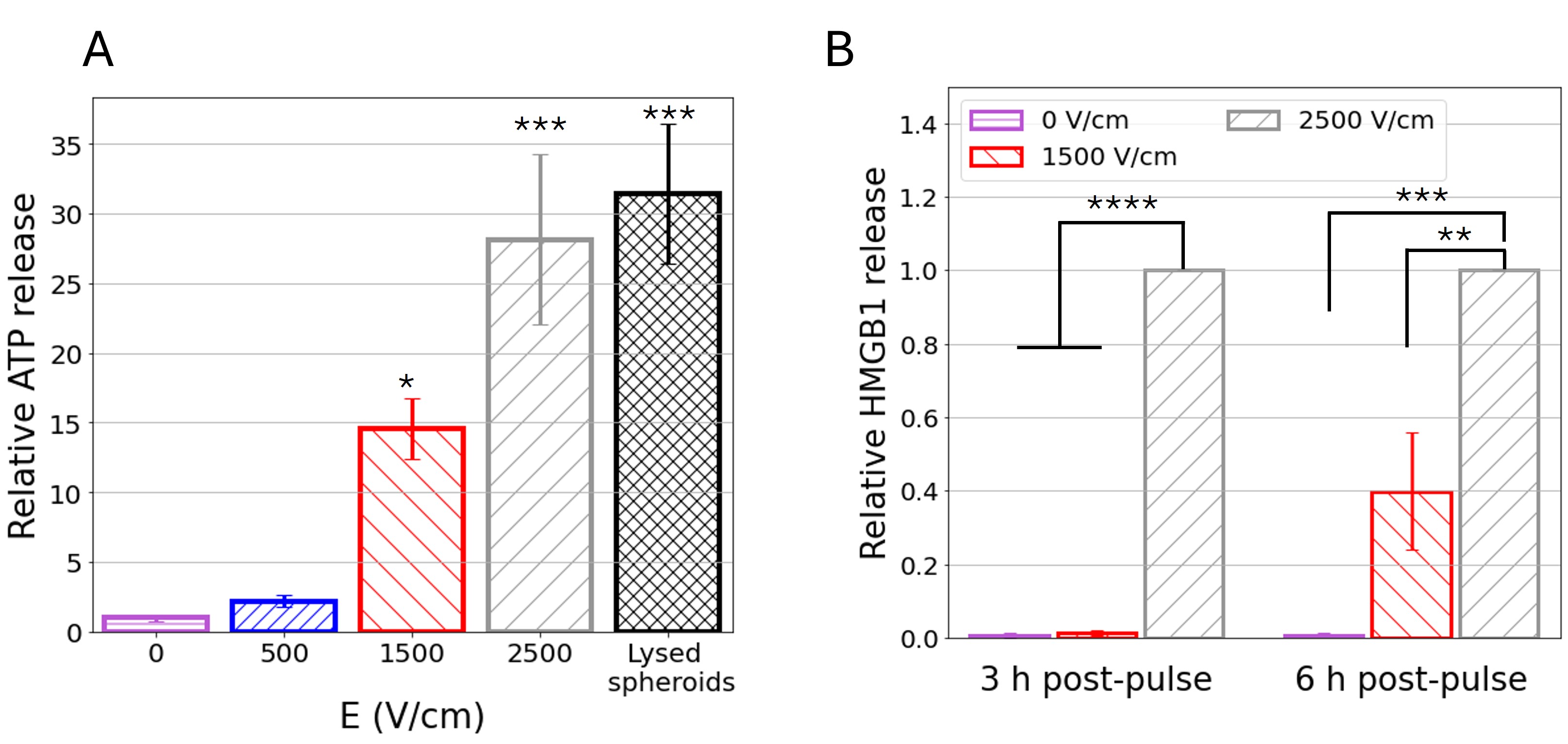}
	\caption{\textbf{DAMP release following electroporation treatment.}  (A) ATP level in supernatant measured 10 minutes after pulse treatment, or after lysis of pools of 3 spheroids (relative to ATP level at 0 V/cm). ATP release luminescence was normalized on luminescence from the controls of each experiment. Statistical significance was assessed using an one-way ANOVA followed by a Dunnett’s multiple comparison test, in comparison to the 0 V/cm condition. * $p < 0.05$, *** $p < 0.001$. Data are shown as the mean ± SEM from three independent experiments (N=3) each condition consisting of three pools of three spheroids, including 9 spheroids per experiment.  (B) HMGB1 level in supernatant measured 3 and 6 hours after pulse treatment of a pool of 27 spheroids (relative to HMGB1 level at 2500 V/cm). Statistical significance was assessed using an one-way ANOVA followed by a Dunnett’s multiple comparison test, in comparison to the 2500 V/cm condition of the same timeframe. ** $p < 0.01$, *** $p < 0.001$, **** $p < 0.0001$. Data are shown as the mean ± SEM from three independent experiments (N=3), each condition consisting of a pool of 27 spheroids.
	} 
	\label{data_tumourgrowth}
\end{figure}

In a clinical setting, cancer treatment-induced cell death might trigger an immune response against residual tumour cells, thereby enhancing treatment efficacy \cite{vacchelli2014trial}. To assess the immunogenicity of the treatment-induced cell death, two DAMPs were assessed in this work: ATP and HMGB1. ATP assays were carried out shortly (10 minutes) after pulse treatment (Figure~\ref{data_tumourgrowth}A) and 24 hours after treatment (Supporting Figure \ref{supplementary}). ATP release amount was proportional to the applied pulse intensity, with almost all cellular ATP released at 2500 V/cm, the highest applied voltage. These results suggest a positive correlation between the electrical field intensity and the ATP release, which is in line with previously reported experimental observations  \cite{polajzer2020analysis}. The experiment assessing extractable intracellular ATP 24 hours post-treatment shows that cells from spheroids treated at high intensity electric fields are not capable of restoring the production of ATP (Supporting Figure \ref{supplementary}).

\begin{figure}[H]
	\centering
	\includegraphics[width=15cm]{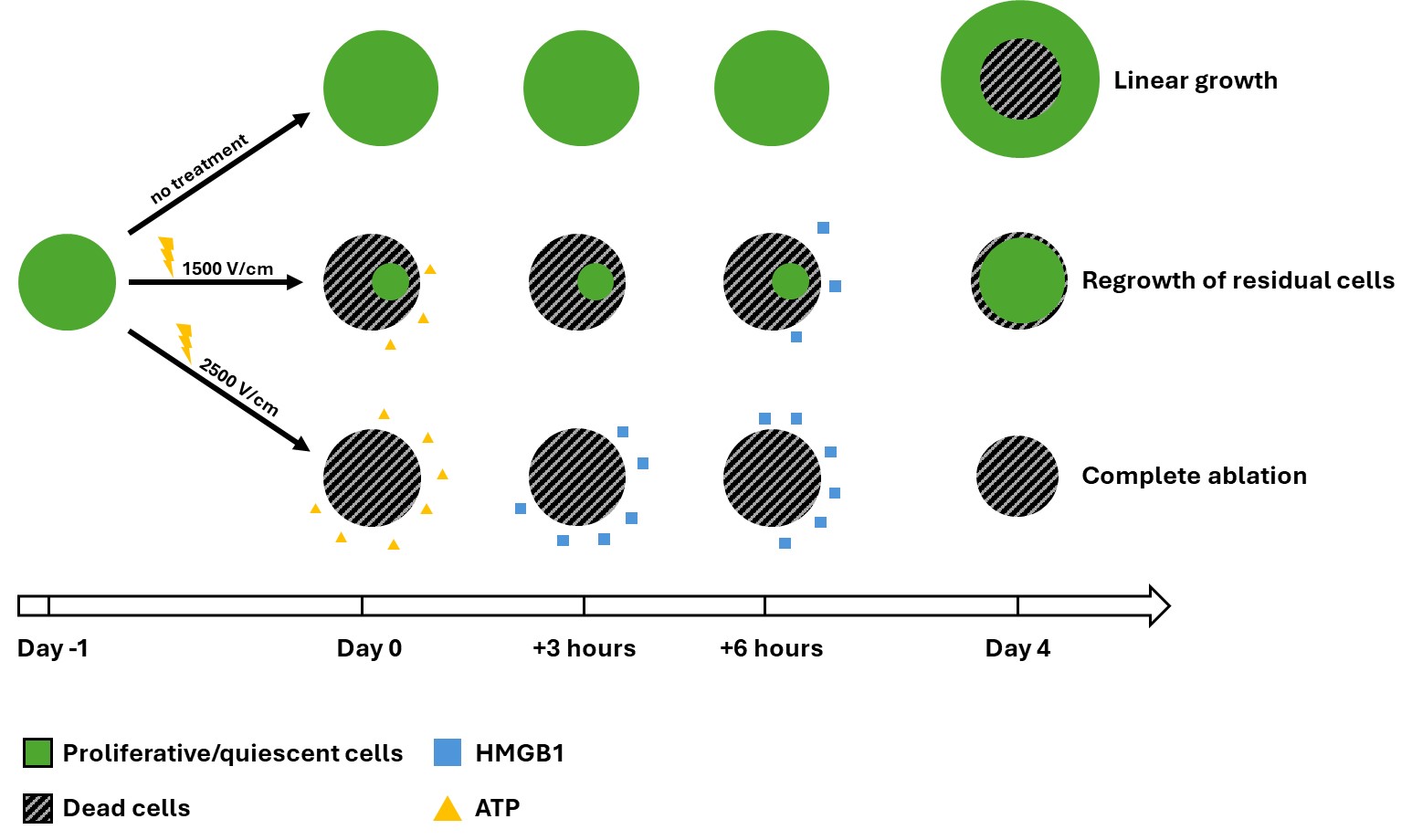}
	\caption{\textbf{Schematic summary of the \textit{in vitro} experiments.} In control (untreated) spheroids, a proliferation gradient was observed over time, characterized by an increasing size of spheroids and the formation of non-viable core in the center. On day 0, treated spheroids underwent pulse application at varying intensities. As a summary of the observations of our experimental study, the dynamics of the control and 500 V/cm, as well as 1500 V/cm and 2500 V/cm treated spheroids are schematized. At 500 V/cm the effects were comparable to the control (0 V/cm), while 1500 V/cm pulse intensity induces an almost complete spheroid ablation, with the survival of residual cells leading to increased spheroid regrowth at later timepoints, in comparison to the control. 2500 V/cm pulse intensity leads to a complete loss of viability in the large majority of treated spheroids. The amount and timepoints of ATP and HMGB1 release vary depending on pulsing conditions.
	} 
	\label{experimental setup}
\end{figure}

The dynamics of HMGB1 release in the supernatant was analysed at 3 and 6 hours after treatment (Figure~\ref{data_tumourgrowth}B). Immunoblotting results showed release of HMGB1 at 3 hours post-treatment only at 2500 V/cm, which was the strongest voltage tested. In contrast, no HMGB1 was detected at 1500 V/cm and lower voltage. Nonetheless, at the 6 hours post-treatment time-point, extracellular HMGB1 was also detected at 1500~V/cm, though in a lower amount than in the 2500~V/cm condition. These results are consistent with the previous observation of apoptosis induction, showing that cell death was more extensive and happened faster in the strongest pulse condition compared to lower pulse intensity. These results indicate that the tested treatment does induce an immunogenic cell death, dependent on the applied electric field intensity. Hence, the heterogeneity of the electric field in clinical IRE treatment will inevitably impact the timing of immunogenic signals. The overall results of this experimental section are summarized in Figure~\ref{experimental setup}. 

\subsubsection{pH measurement }
Under our experimental conditions, the applied protocols did not result in pH alteration. As such, electrolysis is unlikely to be the reason for the observed cell death. The results are represented in Table \ref{pH_table}. 
\begin{table}[h]
	\centering
	\caption{pH in pulsed electroporation buffer (ZAP) or buffer and water combination.}
	\begin{tabular}{p{5cm} p{5cm} p{5cm}}
		\hline
		Applied pulse intensity (V/cm) &
		pH in 150~\(\mu\)L ZAP &
		pH in 100~\(\mu\)L ZAP (+ 50~\(\mu\)L H\(_2\)O) \\
		\hline
		0    & 7.53 $\pm$ 0.08 & 7.62 $\pm$ 0.05 \\
		2000 & 7.51 $\pm$ 0.03 & 7.60 $\pm$ 0.02 \\
		2500 & 7.59 $\pm$ 0.02 & 7.69 $\pm$ 0.02 \\
		\hline
	\end{tabular}
		\label{pH_table}
\end{table}
\subsection{Numerical results}
\label{Numerical_results}
We now present the main numerical results of the mathematical model, which serves as a digital twin of the experimental set-up. 
As already stated by H. Byrne and co-authors, the bidimensional setting enables fast computation and captures key biological features \cite{bull2020mathematical}, so we focus now on this bidimensional setting even though our modelling is not restrictive to it.

 In the numerical simulations, different groups are denoted EP0, EP500, EP1000, EP1500, EP2000 and EP2500, respectively, which corresponds to measures obtained in groups exposed to 0 V/cm (control), 500 V/cm, 1000 V/cm, 1500 V/cm, 2000 V/cm or 2500 V/cm, respectively. Among listed conditions, biological data obtained with 0 V/cm, 500 V/cm, 1500 V/cm, and 2500 V/cm were used to calibrate the mathematical model, while 1000 V/cm and 2000 V/cm were used to validate it. In addition, to confirm the findings of our mathematical model, we conducted retrospective micrograph analysis which provided insight into numerically proposed mechanisms of spheroid regrowth. 
\subsubsection{Set-up of numerical simulations}
The hybrid IBM is parametrised using parameter values retrieved from the literature, wherever possible. Details of model parametrisation are given in Appendix \ref{calibration}, along with the full list of parameter values and related references (\textit{cf.} Table~\ref{ch4:table2}).
For the numerical simulations we report on, we use the 2D spatial domain $\Omega:=[-1,1]^2$. Under the parameter choice of Table~\ref{ch4:table2}, this is
equivalent to considering a square region of a 2D cross-section of the \textit{in vitro }culture assay of area $4 \,mm^2$. Furthermore, to carry out numerical simulations of the  model, we use the space-step $\chi= 0.033$~{\it mm} and the time-step $\tau = 0.005$~{\it days}. Finally, unless otherwise specified, we carry out numerical simulations for 5 days (which we
count from day -1 to day 4). All simulations are performed in \textsc{Python} using the scientific computing libraries Numpy and Scipy \cite{Harris2020array,2020SciPy-NMeth} and the graphics library Matplotlib \cite{Hunter:2007}.

Given the stochastic nature of the model, all the results we present in this section are obtained by averaging over 3 simulations. Note that we did not perform a parameter estimation procedure, but a calibration of our model based on experimental data. A sensitivity analysis of our
model to some free parameters of interest is presented in Appendix \ref{sensitivity analysis}. It should also be noted that the standard deviation
between the 3 simulations is relatively small which allows not to increase the number of
simulations (and their relative computational cost). 

At the initial time of simulations, a certain number of
tumour cells is placed in the domain, tightly packed in a circular
configuration positioned at the centre of the domain, reproducing
the geometry of a spheroid. We assume that at the initial time of simulations, tumour cells are in a proliferative state, which means that
\begin{equation}
	n_{ij}^0=p_{ij}^0=900\exp[-65(i\chi-x_{11}^*)^2-65(j\chi-y_{11}^*)^2],
	\label{eq:init_tum}
\end{equation}
with $(x_{11}^*,y_{11}^*)=(0,0)$. Such initial conditions are chosen to ensure that the initial simulated tumour area aligns with the initial spheroid area computed in the experiments. 
The initial concentration of nutrients is set to $C_{\text{out}}$ everywhere in the domain.
\subsubsection{Control scenario: spheroid growth without electroporation effects}
We first establish a preliminary scenario where tumour cells  proliferate and transition from proliferative to quiescent and necrotic state according
to the rules described in Section \ref{growthPriorElect}, when electroporation is not applied.

\begin{figure}[H]
	\centering
	\includegraphics[width=16cm]{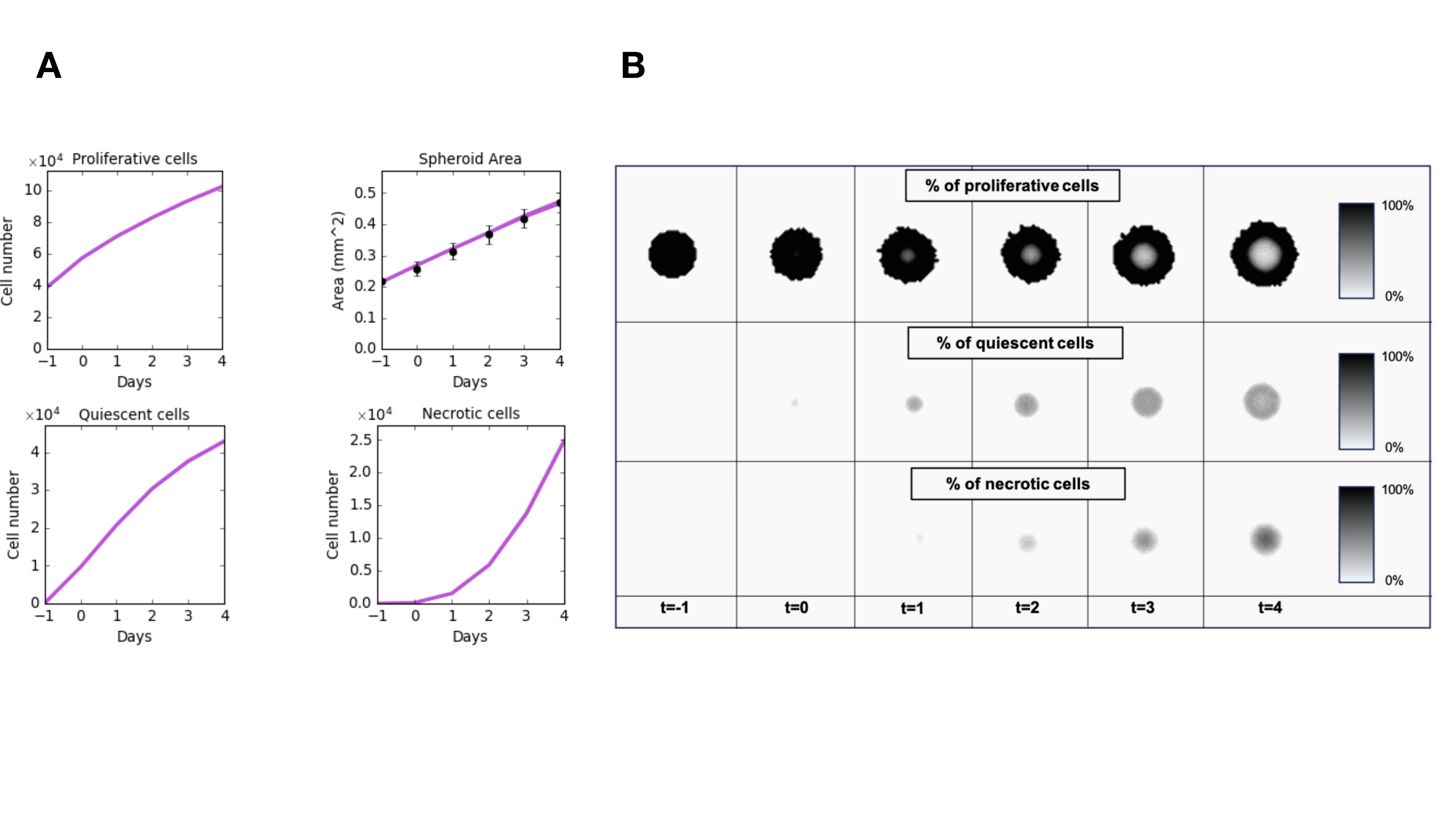}
	\caption{\textbf{Control scenario: tumour growth without electroporation effects.} (A) Time evolution of the proliferative, quiescent and necrotic cell number, as well as the time evolution of the spheroid area in control conditions. These results correspond to the average and $+/-$ standard deviation over 3 simulations. The black markers highlight average (scatter points) and standard deviation
		(error bars) of the experimental data that are used to carry out model calibration.   (B) Example of
		the spatial distribution of proliferative, quiescent and necrotic cells at different times of the simulation. Black (resp white) = 100
		(resp. 0) \% of proliferative/quiescent/necrotic cells.} 
		\label{controlScenario}
\end{figure}

  The plots in Figure~\ref{controlScenario}A show the time evolution of the number of proliferative, quiescent and necrotic cells of group EP0 (\textit{i.e.} the control group), as well as the time evolution of the spheroid area. Figure~\ref{controlScenario}B display samples of the spatial distributions of proliferative, quiescent and necrotic cells at different times of one simulation. As shown by Figure~\ref{controlScenario}A, the number of proliferative cells increases over time, although the speed of growth tends to decrease due to cells transitioning to quiescent or necrotic states. Quiescent cells exhibit a similar trend, with a rapid increase in their numbers from Day -1 to Day 3, followed by a slower rise after Day 3. This behaviour is probably due to the gradual depletion of nutrients in the central part of the domain and the transition of quiescent cells to necrotic states. Necrotic cells begin to appear between Day 0 and Day 1, and their number increases until the end of numerical simulations. Finally, the area of the spheroids increases over time, and its dynamics is similar to the one observed experimentally in Figure \ref{damps_release}B.
   
  These simulations allowed us to calibrate the model parameters related to tumour cells to qualitatively reproduce the growth of the spheroids area in control conditions observed in the experiments. The
  other simulations were carried out keeping the values of these parameters fixed and equal to
  those used for these simulations.

\subsubsection{Electroporation dynamics and release of ATP}
We now explore the impact of electroporation on tumour cell death and regrowth, as well as on ATP and HMGB1 release.  For these simulations,
the initial composition of tumour cells corresponds to that obtained at day 0, along with the corresponding configuration of proliferative, quiescent, and necrotic cells. We also assume that electroporation is applied with~500 V/cm, 1500 V/cm, and 2500 V/cm voltage intensities.

\begin{figure}[H]
	\centering
	\includegraphics[width=14cm]{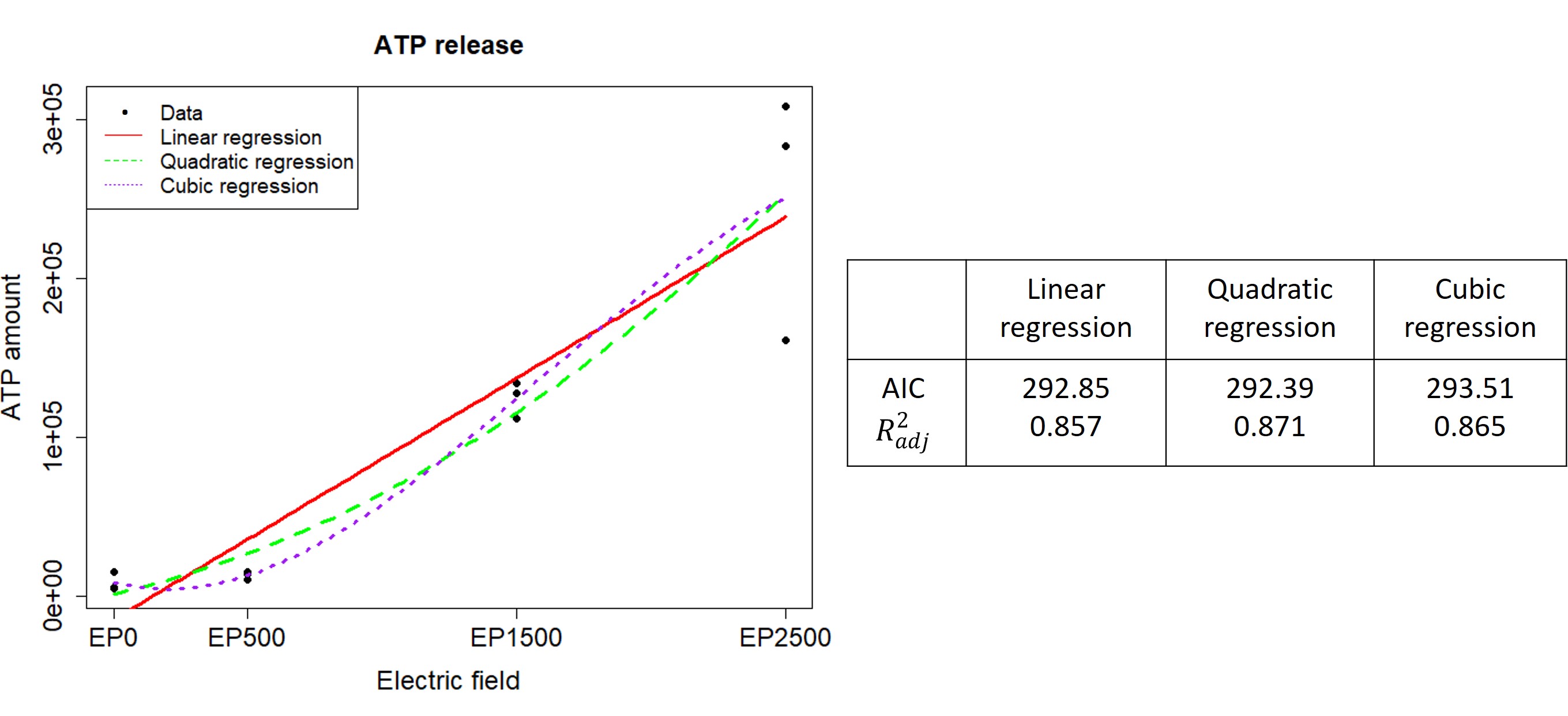}
	\caption{\textbf{Modelling the relationship between intensity of the electric field and  amount of ATP released in the supernatant. }
		Left panel: Comparison between linear, quadratic and cubic regression to investigate the relationship between the magnitude of the electric field $\bold{E}$ and amount of ATP in the supernatant 10 minutes after electroporation. Right panel: corresponding AIC and $R^2$ adjusted obtained from  linear, quadratic and cubic regression.} 
		\label{ATPrelease}
\end{figure}

To study how the strength of the electric field affects ATP release from the cells, we used the experimental data in Figure~\ref{data_tumourgrowth}A, showing ATP levels 10 minutes after electroporation, along with the corresponding electric field intensity $\bold{E}$ used in the experiments. Using polynomial regression in \faRProject with the \texttt{lm()} function \cite{Rsoft}, we compared linear, quadratic, and cubic models, focusing on the AIC and adjusted $R^2$ metrics. Results are shown in Figure \ref{ATPrelease}. Although all three models fit the data well, the quadratic model provided the best fit, with the lowest AIC (292.39) and highest adjusted $R^2$ (0.871), indicating a good balance between accuracy and simplicity.


	
\subsubsection{Death of cells and release of HMGB1}
We now assume that cancer cell death resulting from electroporation is proportional to the amount of ATP released in the supernatant. Following the experimental results shown in Figure~\ref{sph_death}D, we also assume that the time before cell death occurs is inversely proportional to the intensity of the electric field applied. With these assumptions in place, we aim to validate the effectiveness of our model in qualitatively replicating the dynamics of spheroid death observed in the experiments. 

\begin{figure}[H]
	\centering
	\includegraphics[width=15cm]{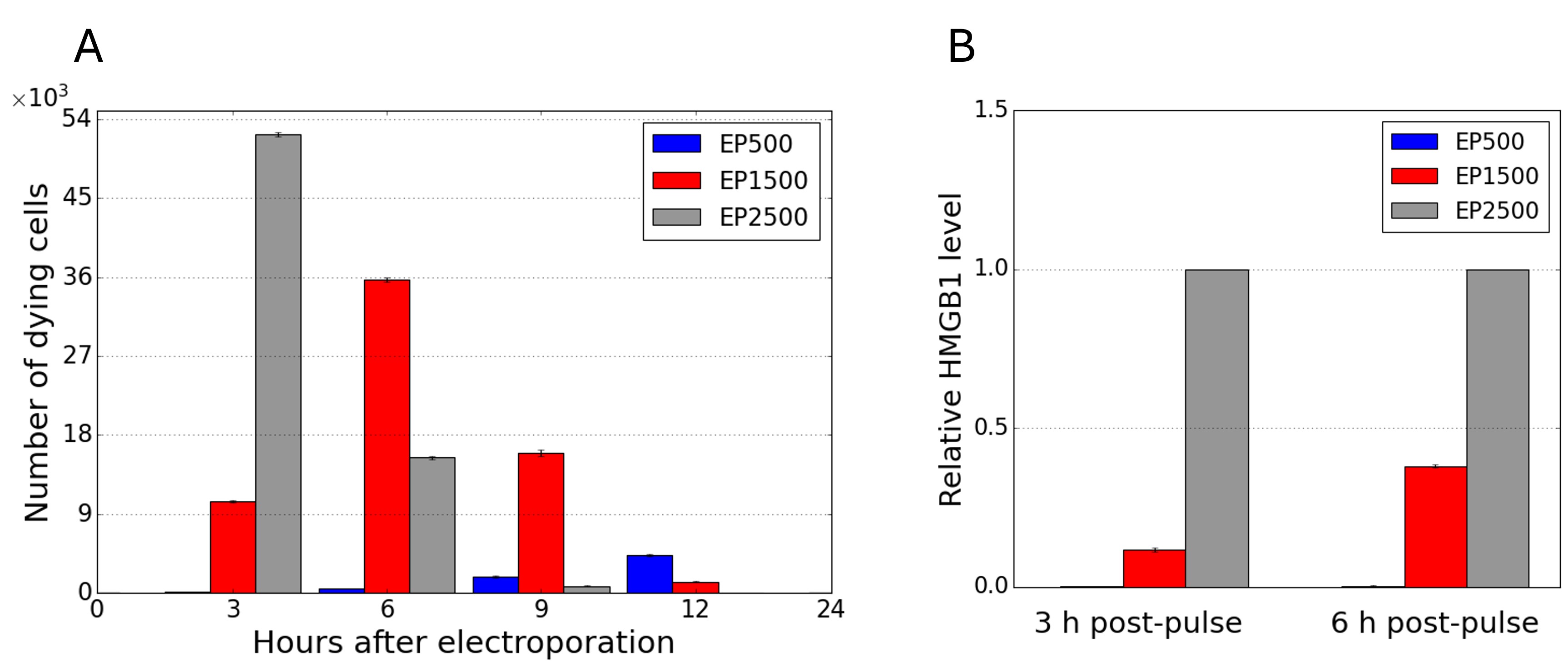}
	\caption{\textbf{Apoptotic cell death and release of HMGB1. }
		(A) Number of cells in a dying state immediately and at various time intervals (3 h, 6 h, 9 h, 12 h, and 24 h) after the pulse exposure. (B) Total amount of HMGB1 in the domain (\textit{cf.} Eq.~\eqref{HMGB1}) (relative to the amount of HMGB1 of group EP2500) 3 and 6 hours after the pulse exposure. The error lines represent the standard
deviation between 3 simulations.} 
	\label{HMGB1release}
\end{figure}

Figure~\ref{HMGB1release}A illustrates the number of cells in a dying state immediately and at various time intervals (3 h, 6 h, 9 h, 12 h, and 24 h) post-electroporation. These results show that, in group EP2500, the highest number of cells in a dying state is reached 3 hours post-electroporation, whereas in groups  EP1500, this peak is reached mostly 6 hours after electroporation. Finally,  cell death of group EP500 is almost absent, as the effect of electroporation is almost absent for such low electric pulse intensity. These numerical results are in line with those obtained experimentally.  

Then, assuming that dead cancer cells release HMGB1 in the domain (\textit{cf.} Eq.~\eqref{HMGB1}), we investigate the dynamics of HMGB1 release produced by our model. Figure~\ref{HMGB1release}B shows the amount of HMGB1 in the domain (relative to the amount of HMGB1 of group EP2500) 3 and 6 hours post electroporation. These results show that, 3 hours post-electroporation, HMGB1 is almost uniquely released by cells in group EP2500. Only a very low amount of HMGB1 is also released by cells  in  group EP1500. Then, 6 hours post-electroporation, the amount of HMGB1 released by cells in group EP1500 slightly increases, but still remains very low compared to that released by cells in  group EP2500.

 Taken together,  these numerical results exhibit a good qualitative agreement with the experiment presented in Section \ref{in vitro_Results}. Moreover,  these findings might shed light on the roles played by dead cancer cells in influencing the release of HMGB1 in the supernatant. According to our model's assumptions, high electric pulses cause  a rapid tumour cell death, leading cells exposed to a high electrical field intensity to release HMGB1 earlier than cells exposed to lower electrical field intensities.
\subsubsection{Spheroid regrowth after electroporation}
The experimental results presented in Section \ref{in vitro_Results} showed a complex interplay between electric pulse intensity and spheroid growth, with lower intensities having minimal impact, moderate intensities stimulating growth, and higher intensities causing significant growth inhibition (\textit{cf.} Figure~\ref{damps_release}B-C). Here we verify the ability of our model to reproduce such
dynamics by exploring the regrowth of tumour cells  over 4 days after electroporation.

\begin{figure}[H]
	\centering
\includegraphics[width=\linewidth]{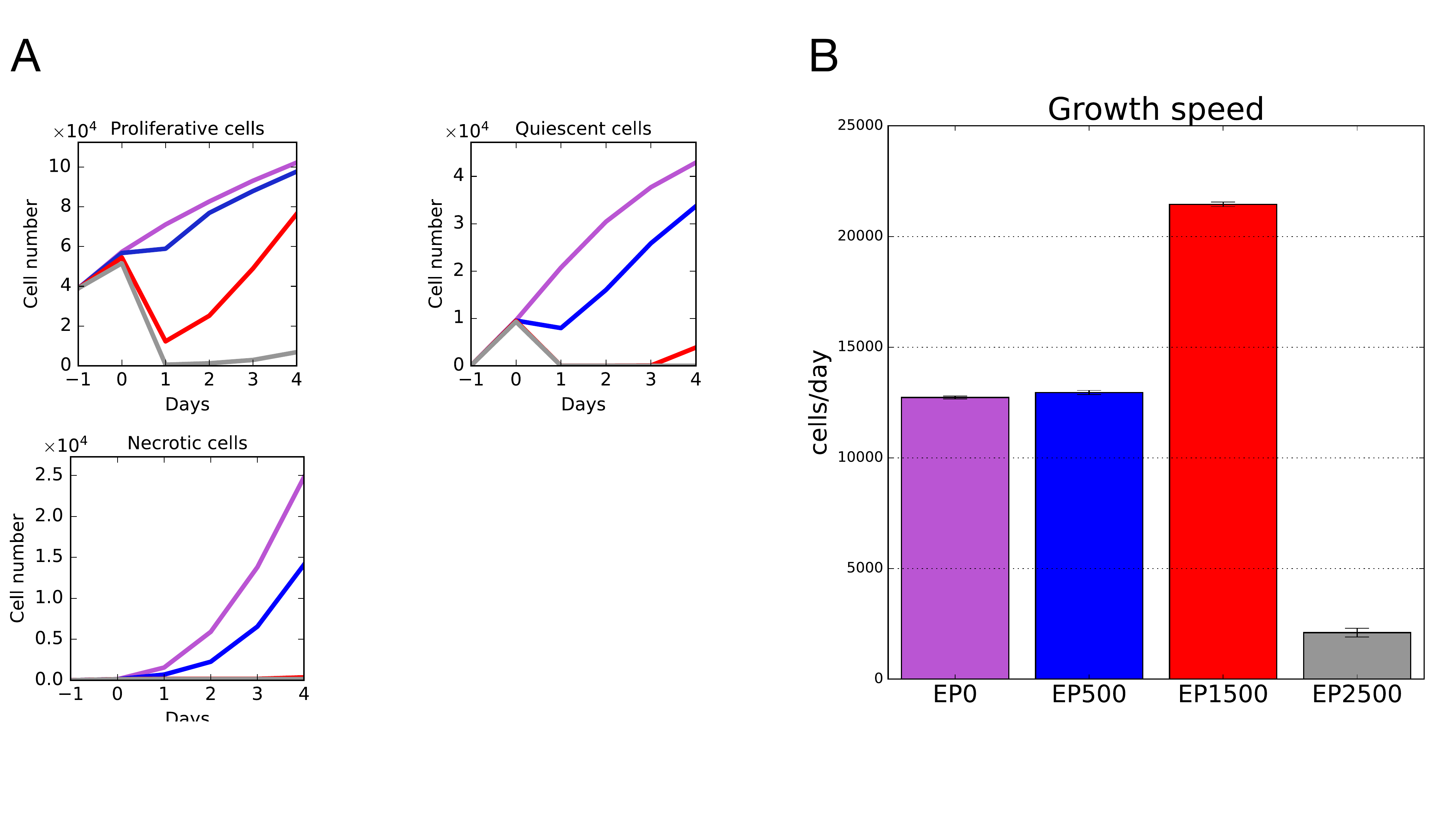}
	\caption{\textbf{Spheroid regrowth after pulsed electric field. }
		Panel A: time evolution of the proliferative, quiescent and necrotic cell number for increasing intensities of the electric field. These results correspond to the average and $+/-$ standard deviation over 3 simulations. Panel B: corresponding speed of growth of proliferative cells between Day 1 and Day 4. The error lines represent the standard
		deviation between 3 simulations.} 
	\label{tumourGrowth}
\end{figure}

The plots in Figure~\ref{tumourGrowth}A show the time evolution of the number of proliferative, quiescent, and necrotic cells, as well as the time evolution of the spheroid area. It is interesting to note that the speed of spheroid area growth between Day 1 and Day 4 appears similar for EP0, EP500, and EP1500. This behaviour arises from the locally low effective compressibility of the cell population (see the tumour cell proliferation rules described in Section 2.2.1), which ensures that the spheroid area expansion is mainly driven by the dynamics of cells at the spheroid periphery. The mathematical model shows that after the electric pulse at 1500V/cm, the proportion of proliferative cells increase in the spheroid, that leads to a increase of growth speed in cell/day as measured in the experiments. Indeed, Figure~\ref{tumourGrowth}B illustrates the speed of growth of proliferative cells between Day 1 and Day 4, which accurately reproduce the experimental data tracking the speed of growth of live cells within the spheroids (\textit{cf.} Figure~\ref{damps_release}C).

\begin{figure}[H]
	\centering
	\includegraphics[width=\linewidth]{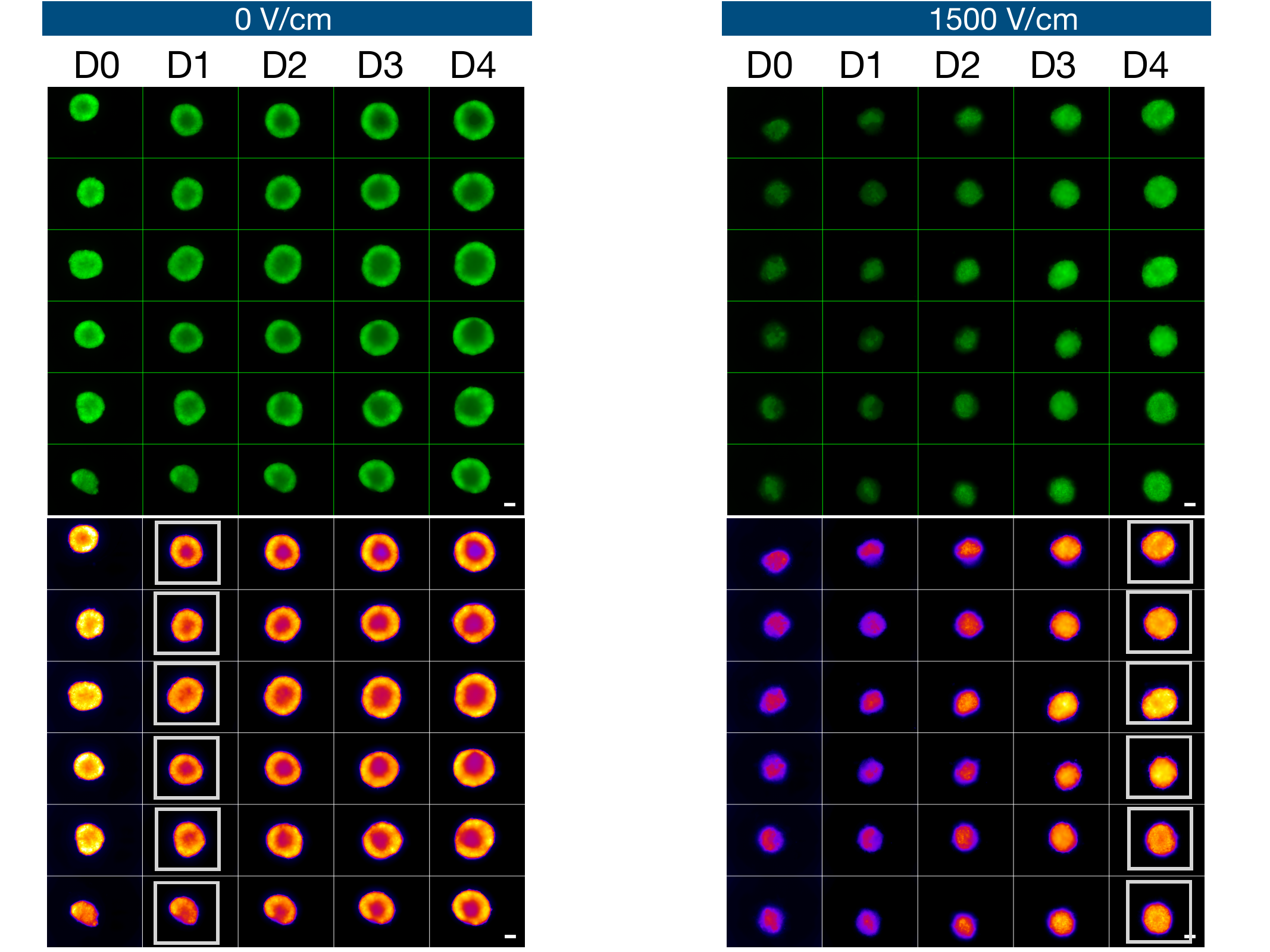}
	\caption{\textbf{Retrospective examination of micrographs showing spatial distribution of GFP fluorescence intensity in control and 1500 V/cm-treated spheroids. }
		Panels display six spheroids (selected from one experiment on a “first-to-come” basis) for 0 V/cm (left) and 1500 V/cm (right), with GFP fluorescence shown in green colour. The bottom panels present the same spheroids rendered in a fire-colour scale representing GFP fluorescence intensity. These micrographs, highly consistent across replicates and independent experiments, corroborate our numerical findings: in the 1500 V/cm condition, the number of necrotic cells does not increase over the four days post-treatment, while proliferative cells continue to expand. White frames in bottom panels delineate spheroids of comparable size, highlighting that although 1500 V/cm-treated spheroids reached the same diameter as controls on Day 1, they did not develop the necrotic core characteristic of untreated spheroids of similar size. Scale bar = 200 \textmu m. } 
	\label{tumourGrowth1}
\end{figure}

Upon the application of electroporation at Day 0, Figure~\ref{tumourGrowth}A shows that, at low electric voltages, electroporation does not have substantial impact on cell viability, as evidenced by the comparable proliferative cell growth between groups EP0 and EP500.  However, when a sufficiently high electric field is applied (\textit{i.e.} 1500 V/cm), there is an immediate decrease in the numbers of proliferative and quiescent cells between Day 0 and Day 1. Then, starting from Day 1, the number of proliferative cells increases again, with the rate of increase varying depending on the intensity of the electric field applied. 
A similar trend is reflected in the dynamics of the spheroid area. It is interesting to note that our numerical results indicate that quiescent and necrotic cells follow different dynamics. In particular, for sufficiently high electric field intensity (\textit{i.e.} 1500 V/cm), few or no new quiescent and necrotic cells emerge after electroporation. 
In order to validate this assumption, we retrospectively reinvestigated the spatial distribution of GFP fluorescence intensity across spheroids (Figure \ref{tumourGrowth1}). The spatial fluorescence analysis corroborated mathematical model results, revealing the absence of a distinct necrotic core in spheroids treated at 1500 V/cm.


\subsubsection{\textit{Post hoc} validation of model predictions with additional experimental data}
The experimental data described in Section \ref{in vitro_Results} were used to calibrate the parameters of the mathematical model, allowing it to replicate the observed results at 0 V/cm, 500 V/cm, 1500 V/cm and 2500 V/cm. To evaluate the predictive capability of the model, we now include experimental data obtained at 1000 V/cm and 2000 V/cm. Using the previously calibrated parameters, numerical simulations of the mathematical model are now  performed considering an electric field intensity of 1000 V/cm and 2000 V/cm, and compared to the corresponding experiment results.

\begin{figure}[H]
	\centering
	\includegraphics[width=14cm]{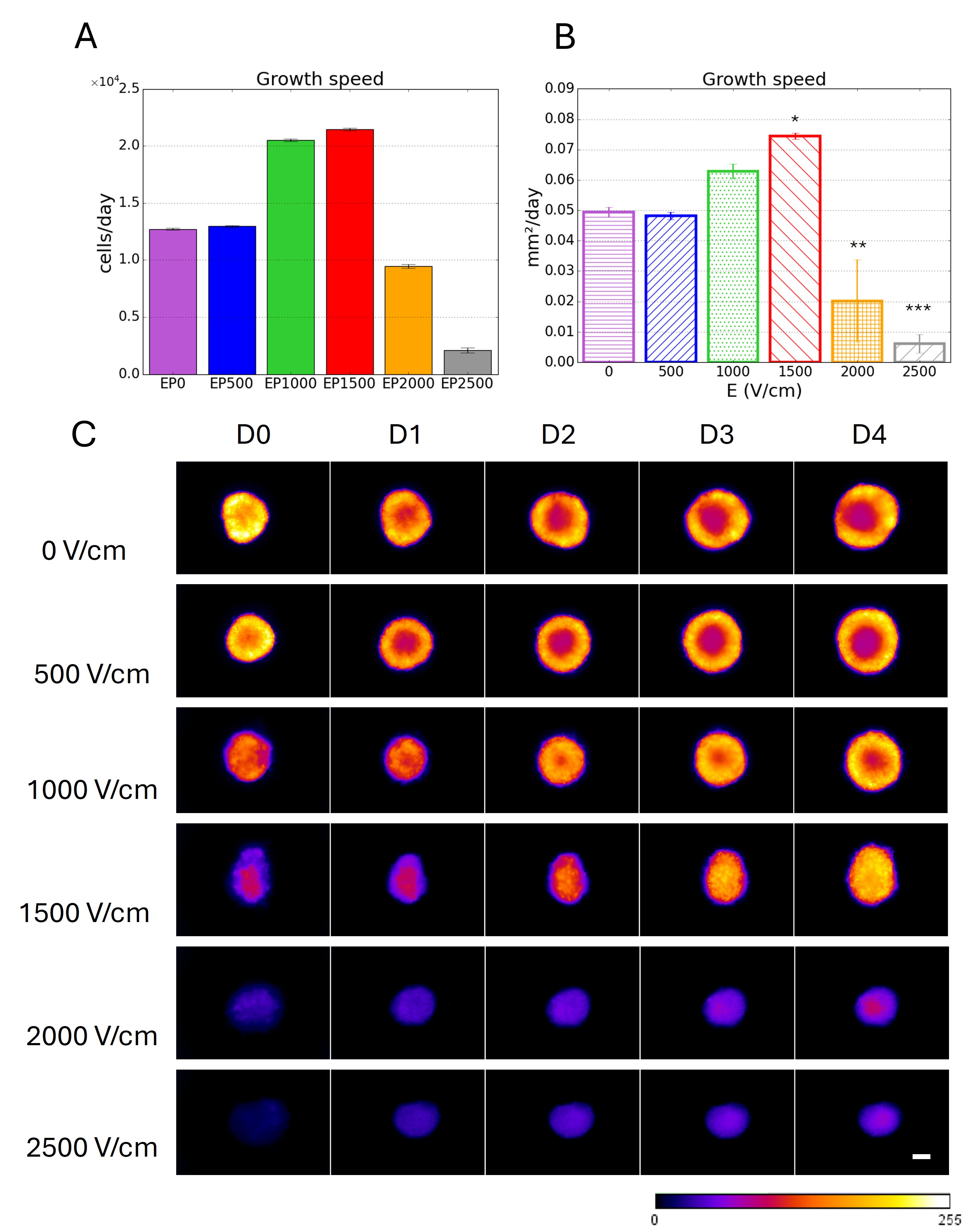}
	\caption{\textbf{Comparison of spheroid growth determined by numerical predictions of the mathematical model and experimental results.}
			(A) Spheroid growth speed predicted by the mathematical model. Data are shown as the mean ± standard deviation from three simulations. (B) Spheroid growth speed based on experimentally determined growth rates (derived from GFP-positive areas). Data are shown as the mean ± SEM from three independent experiments comprising at least 6 spheroids. Statistical significance was assessed using one-way ANOVA followed by a Dunnett’s multiple comparison test, in comparison to the 0 V/cm condition.  * $p < 0.05$,** $p < 0.01$, *** $p < 0.001$. (Scale bar = 200 µm). (C) Representative micrographs showing spatial distribution of GFP-positive spheroid zones represented in a fire-colour scale reflecting GFP fluorescence intensity.}
	\label{modPrediction}
\end{figure}

First, the growth speed of the spheroids after treatment is determined (Figure~\ref{modPrediction}A). The model predicts faster growth in the 1000 V/cm group (EP1000) compared to the control group (EP0) and a slower growth in the 2000 V/cm group (EP2000). This result is in agreement with the outcome of additional experiment results (Figure \ref{modPrediction}B). Interestingly, we could also determine that the necrotic core diminishes with increasing field intensity (Figure \ref{modPrediction}C).

\begin{figure}[H]
	\centering
	\includegraphics[width=15cm]{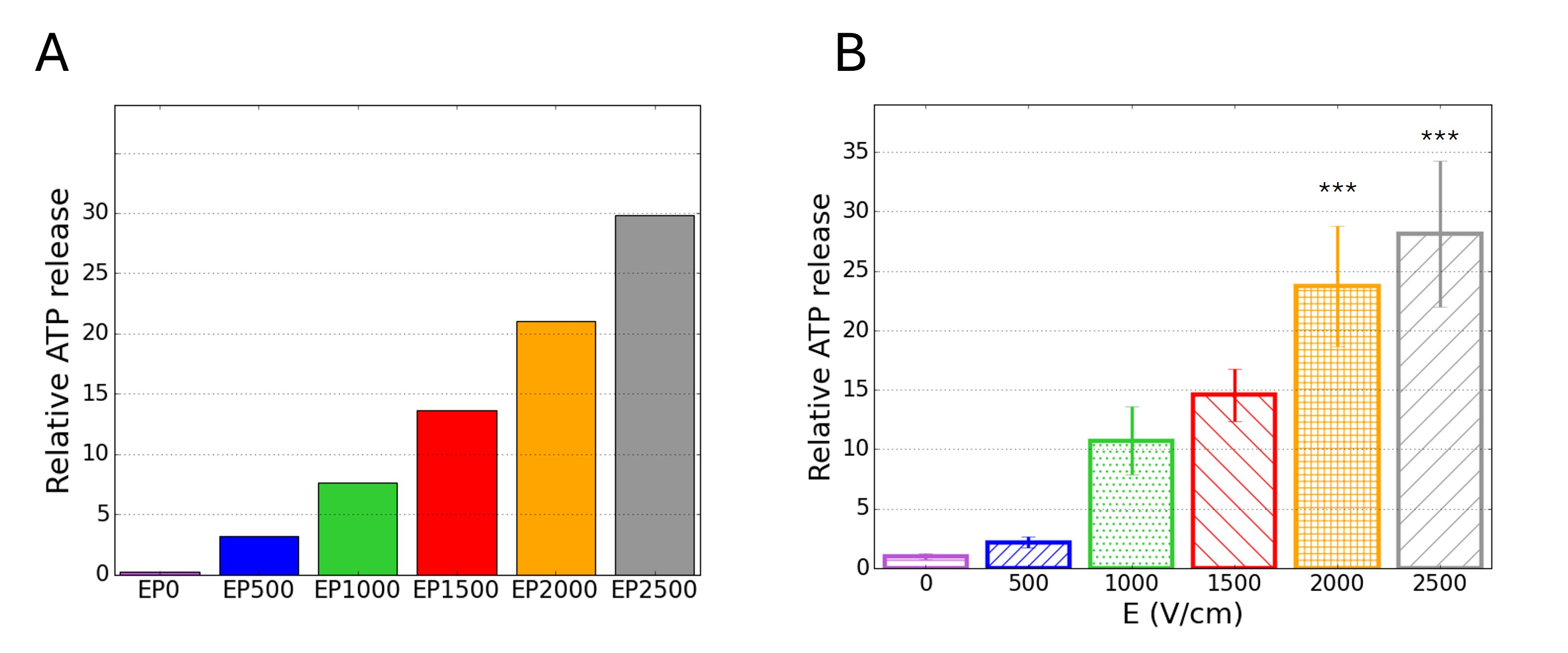}
	\caption{\textbf{Comparison of ATP release determined by numerical predictions of the mathematical model and experimental results.}
			(A) Results obtained by numerical predictions of the mathematical model.  (B) Data determined in experimental results. Data are shown as the mean ± SEM of the mean from three independent experiments comprising at three pools of three spheroids. Statistical significance was assessed using one-way ANOVA followed by a Dunnett’s multiple comparison test, in comparison to the 0 V/cm condition.  *** $p < 0.001$.}
	\label{modPredictionATP}
\end{figure}

Subsequently, we compare the levels of released ATP predicted by the mathematical model and those obtained in the experiments (Figure \ref{modPredictionATP}). The quadratic regression model, employed to describe the relationship between electric field intensity and ATP levels in the supernatant, accurately predicts ATP amounts at 1000 V/cm and 2000 V/cm (Figure \ref{modPredictionATP}).

\begin{figure}[H]
	\centering
	\includegraphics[width=16cm]{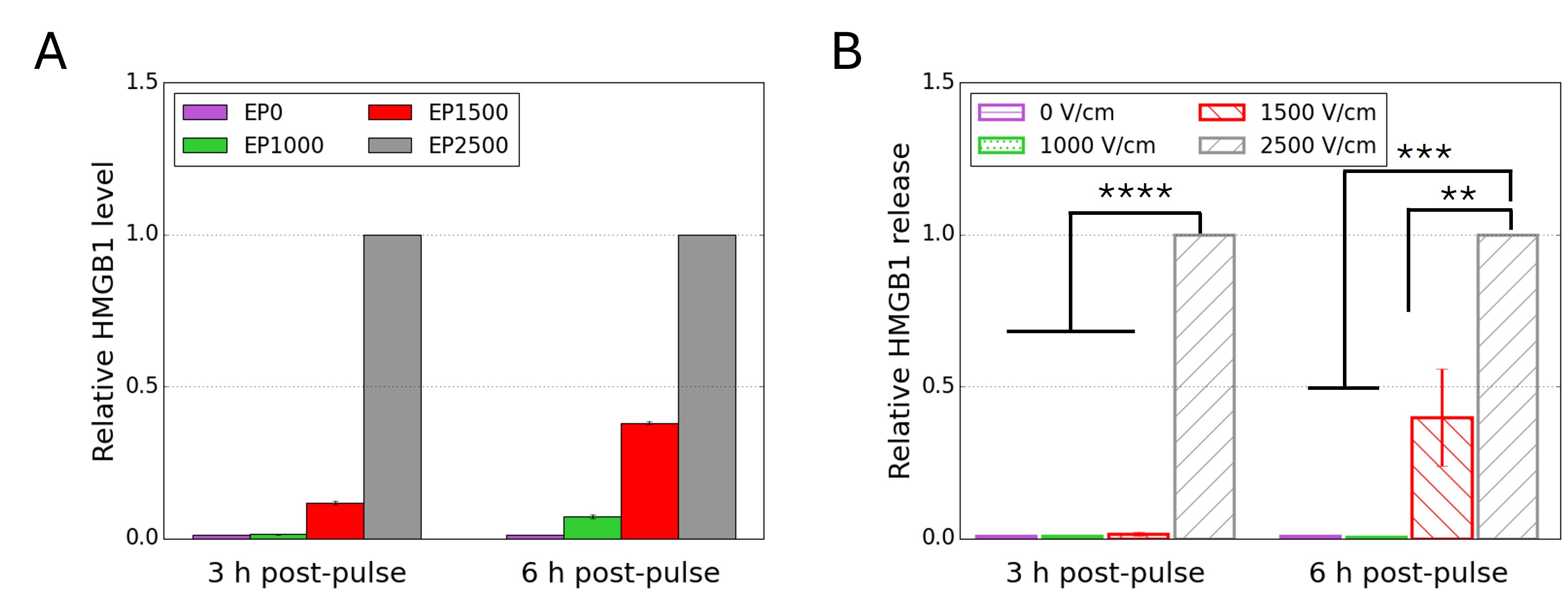}
	\caption{\textbf{Comparison of HMGB1 release determined by numerical predictions of the mathematical model and experimental results. }
		(A) Total amount of HMGB1 in the domain (cf. \eqref{HMGB1}) (relative to the amount of HMGB1 of group EP2500) 3 and 6 hours post electroporation; data are shown as mean ± standard deviation from three simulations. (B) HMGB1 levels in the supernatant measured 3 and 6 hours after electroporation (relative to HMGB1 levels at 2500 V/cm). Statistical significance was assessed using one-way ANOVA followed by a Dunnett’s multiple comparison test, in comparison to the 2500 V/cm condition of the same timeframe. ** $p < 0.001$, **** $p < 0.0001$. Data are shown as the mean ± SEM from three independent experiments (N=3), each condition consisting of a pool of 27 spheroids.}
	\label{HMGB1pred}
\end{figure}

We then compared the level of HMGB1 release  predicted by the numerical simulations with those obtained in experimental data (Figure \ref{HMGB1pred}). For an electric field intensity of 1000 V/cm, the model predicts minimal HMGB1 release at 3- and 6-hours post-electroporation, consistent with the experimental observations. At 1500 V/cm, the model slightly overestimates HMGB1 release at  3 hours compared to experimental results. At 6 hours post pulsation the model reflects well the experimental results. Since HMGB1 release at 2000 V/cm was not experimentally measured, we do not report the corresponding numerical simulation results. 

\begin{figure}[H]
	\centering
	\includegraphics[width=16cm]{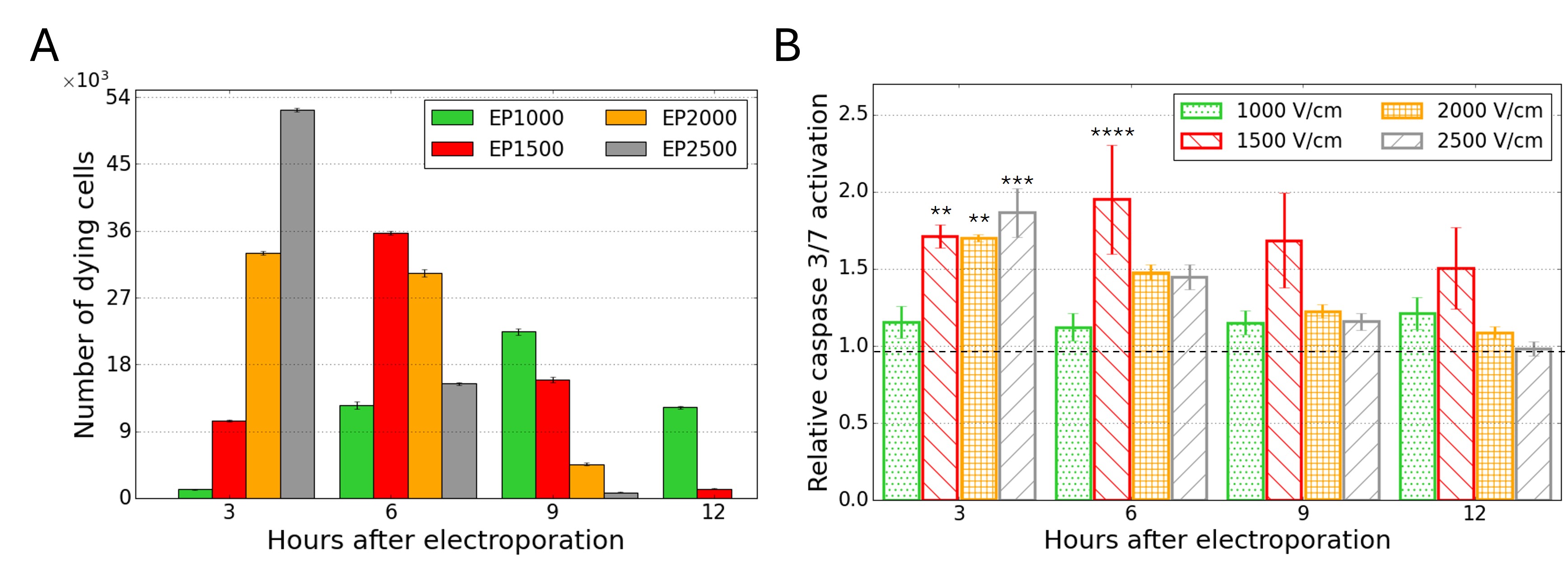}
	\caption{\textbf{Comparison of the kinetics of cell death occurrence following electroporation determined by numerical predictions of the mathematical model and experimental results.}
		(A) Number of cells in a dying state predicted by the mathematical model. Data are shown as the mean ± standard deviation from three simulations. (B) Apoptosis induction determined experimentally based on caspase 3/7 activation, determined from fluorescence intensity measurements and normalized to the 0 V/cm condition. The dashed line in (B) represents the baseline caspase-3/7 activity measured in the untreated control spheroids (0 V/cm), which reflects the constitutive activity detected in the necrotic core. Data are shown as the mean ± SEM  from three independent experiments comprising a pool of 27 spheroids per condition. Statistical significance was assessed using a two-way ANOVA followed by a Bonferroni multiple comparison test, in comparison to the 0 V/cm condition. ** $p < 0.01$, *** $p < 0.001$, **** $p < 0.0001$. }
	\label{celldeath}
\end{figure}

Finally, we modelled the number of dying cells in a 3 hours interval for the first 12h of treatment (Figure~\ref{celldeath}A) and compared them with additional experimental data investigating caspase 3/7 activation (Figure~\ref{celldeath}B). While in the mathematical model we could determine the number of cells that died after electroporation (regardless of cell death type), the biological experiments allowed us to determine the proportion of cells died via caspase 3/7 activation, indicating strong (but not exclusive) evidence of apoptosis.

Both results of Figure~\ref{celldeath} show a similar temporal trend, with the number of dying cells peaking earlier at the highest electric field intensity. In the simulations, maximal cell death occurs at 3 hours for 2500 V/cm and at 6 hours for 1500 V/cm, consistent with experimental observations where caspase 3/7 activity also peaks earlier under stronger electric fields. This agreement supports the model’s ability to capture the kinetics of cell death onset following electroporation. Minor discrepancies between simulated and experimental amplitudes likely arise because the model integrates all death modalities (apoptotic and non-apoptotic) and expresses them as the total number of dying cells, whereas the caspase-3/7 assay principally reports apoptotic events based on fluorescence intensity within caspase-activated regions. These two readouts—quantitative cell counts versus fluorescence-based areas—do not offer equivalent metrics, their concordant timing is sufficient to validate the temporal trends predicted by the model.

\section{Discussion}\label{discussion}
Although IRE has shown encouraging clinical outcomes, tumour relapse still occurs in a subset of treated patients. Recurrence is often attributed to heterogeneous electric field distribution within the tumour, which can leave residual viable cells in underexposed regions. Improving the IRE treatment would ideally require a multifaceted strategy. First, whenever possible, a precise pulse delivery should ensure uniform electric field coverage to achieve complete ablation of the primary tumour. Second, the immune system should be implicated to address surviving or escaping cells. Finally, close clinical monitoring should be used to enable detection and management of relapse at its earliest stage.

Addressing these challenges requires a deeper mechanistic understanding of how electric field parameters shape tumour response and immunogenic signalling. To this end, we developed an experimental set-up coupled with a hybrid IBM to explore cell ablation, DAMP release, and post-treatment regrowth dynamics.

The experimental work focused on 3D spheroids of murine hepatocarcinoma-derived cells, which were exposed to a range of pulse intensities to characterize their biological response. Indeed, the spheroid model does not capture the vascular effects. Nevertheless, the present study methodologically focuses on cell-intrinsic and microtissue-level mechanisms, namely electroporation-induced cell death, DAMP release kinetics, and regrowth dynamics under partial ablation conditions. These processes occur upstream of vascular and immune interactions and are expected to remain relevant \textit{in vivo}, where they are further shaped by perfusion and immune responses. Overall, our experimental results showed a complex interplay between electric pulse intensity, DAMP release, and spheroid regrowth. Specifically, lower intensities had minimal effects on spheroid regrowth, intermediate intensities stimulated regrowth, and highest intensities resulted in substantial regrowth inhibition. The regrowth of spheroids after intermediate intensity treatment has been observed before on a different spheroid model using human colorectal carcinoma cell-derived spheroids \cite{collin2022spatial}, and, together with the findings of this study, warn about the potential of suboptimal IRE to stimulate cancer cell growth.

This is especially relevant for clinical treatments, as pulse intensity covering the tumour can be compromised by suboptimal electrode positioning (which can, for example, be imposed by anatomical barriers) and tumour tissue heterogeneity (\textit{e.g.} adjacent fat) \cite{edd2007mathematical,golberg2015tissue,mathy2020impact}. Eventually, one-third of liver cancer patients treated with IRE shows a progression of the disease one year following the treatment \cite{gupta2021efficacy}. Addressing tumour regrowth and identifying means to prevent it are thus of critical importance. As electroporation, including IRE, can promote the activation of the local and systemic immune response via DAMPs, it is reasonable to combine it with immunotherapies \cite{dai2021irreversible,golberg2015tissue,geboers2021irreversible,zhou2019immunogenic}. DAMPs released after immunogenic cell death trigger a pro-inflammatory milieu, facilitating immune cell maturation and priming cytotoxic T cells. However, the kinetics of cell death and thus the optimal therapeutic scheduling for maximal immune activation may occur over a specific time-limited window.

In this study, we analyzed spheroid viability following IRE pulse exposure, with particular emphasis on the induction of apoptosis. Although apoptosis is usually considered non-immunogenic, growing evidence shows that, under certain conditions, it can trigger an immunogenic response \cite{tesniere2008molecular}. Increasing evidence indicates that IRE induces multiple cell death pathways, each unfolding with distinct kinetics and immunogenic potential \cite{zhang2018molecular}. This may reflect the distinct mechanisms through which IRE induces cell death: while some cells undergo immediate death due to irreversible membrane disruption, others die later as a consequence of sustained loss of intracellular homeostasis despite membrane resealing \cite{rajeckaite2018calcein}. Consequently, the timing of DAMP release peaks may vary. It is thus important to evaluate DAMP profiles over time, for optimal administration of immune checkpoint inhibitors, cytokine therapies, or adoptive T-cell transfer, to enhance anti-tumour immunity while minimizing immune evasion mechanisms. Indeed, optimizing immunotherapy scheduling has become increasingly important, as it can significantly impact treatment effectiveness. By integrating DAMP kinetics and subsequent immune system activation with immunotherapy administration timing, it might be possible to synergistically amplify the systemic immune response, preventing tumour recurrence and improving long-term treatment efficacy \cite{yao2021impact,huo2024optimal,beyranvand2017importance, zemek2024exploiting}. Considering the potential implication of DAMPs, two of them, ATP and HMGB1, were assessed in this work, focusing on the kinetics of their release.

\textit{in vivo}, ATP binds to purinergic receptors and serves as a chemoattractant, as well as inductor of NLRP3 inflammasome activation \cite{fucikova2015prognostic, zhou2019immunogenic, krysko2012immunogenic}. HMGB1, depending on its redox state, binds to different immune receptors (TLR2, TLR4, RAGE, CXCR4, ...), and serves multiple roles, such as chemoattractant or inductor of cytokine production and inflammation \cite{fucikova2015prognostic,he2021role,zhou2019immunogenic, krysko2012immunogenic}. In this work, the release of both directly correlated to the applied pulse intensity. ATP release was assessed shortly after treatment, as it is well-known that electroporation induces immediate ATP release. Later time points are less relevant, as previous studies and our own data (Supporting Figure \ref{supplementary})  indicate that ATP levels decline at later stages \cite{polajzer2020analysis, rols1990electropermeabilization}. The observed correlation between the level of ATP detected in the supernatant and voltage intensity is not surprising, as such results were described with both reversible and irreversible electroporation experiments \textit{in vitro} \cite{jakstys2020correlation, polajzer2020analysis,polajvzer2023immunogenic}.

The release of the second DAMP, HMGB1, was also observed after high intensity pulses, which is in agreement with previous reports \cite{he2021role,dai2021irreversible,zhao2019irreversible}. Extending beyond these reports, our findings demonstrate that HMGB1 release is time-dependent, with intermediate electric field intensity (1500 V/cm) eliciting a delayed response. To the best of our knowledge, this delayed pattern of HMGB1 release has not been previously reported \textit{in vitro} irreversible electroporation studies. Most existing investigations relied on simple 2D cell cultures, which lack the layered organization of proliferative, quiescent, and necrotic zones characteristic of 3D spheroid models. In our study, the delayed HMGB1 release observed at intermediate pulse intensity (1500 V/cm) closely paralleled the postponed peak of apoptosis under the same conditions. The biological experiments performed herein confirm the presence of DAMPs, without directly assessing the immunogenicity of their exposure. Future studies should therefore extend this analysis to \textit{in vitro} and \textit{in vivo} studies involving immune cells. Moreover, future work should expand this analysis to additional DAMPs (\textit{e.g.} calreticulin) to provide a more comprehensive understanding of the immunogenic landscape following IRE.

Beyond DAMPs, high-intensity pulsed electric fields are also known to trigger intracellular ROS generation via direct electrochemical mechanisms (electrolysis) and also indirect cellular responses, such as mitochondrial disruption and activation of NADPH oxidases. ROS generation is considered partially responsible for the plasma membrane permeabilization via the peroxidation of its lipids \cite{gabriel1994generation,breton2018investigation, rems2019contribution,szlasa2020oxidative}. The oxidative stress can trigger different types of cell deaths, including apoptosis, necrosis, autophagy and ferroptosis \cite{valencia2004reactive,circu2010reactive,filomeni2015oxidative,villalpando2021reactive}, and can participate in the activation of the inflammasome, a hallmark of pyroptosis \cite{harijith2014reactive}. Additionally, ROS modulate DAMP signalling. Endoplasmic reticulum stress promotes the exposure of DAMPs, such as the translocation of calreticulin to the plasma membrane. It can also enhance ATP release through the opening of connexin hemichannels, while HMGB1 oxidation modulates its immunological activity \cite{fucikova2015prognostic,panaretakis2009mechanisms,dosch2018mechanisms}. Thus, while our study focused on ATP and HMGB1 as initial ICD markers, the contribution of ROS to these pathways should also be evaluated and is planned for future work, to clarify their relationship with field strength, cell fate, and DAMP dynamics. A further extension could also rely on enriching the model with immune cells \cite{dai2021irreversible,shao2019engineering,zhao2019irreversible}.  It is probable that the addition of such cells would improve the efficiency of the treatment. This might help guide and inspire the design of future clinical trials investigating the potential synergy of electroporation and immunotherapy in cancer treatment \cite{burbach2021irreversible,justesen2022electroporation}.

Partial differential equations coupled to stochastic IBMs capture spatially and temporally heterogeneous tumour dynamics. In this context, macroscopic fields refer to continuous variables—such as nutrient, oxygen, mechanical stress, or electric field distributions—that evolve smoothly in space and time and act collectively on many cells. Meanwhile, individual cells behave stochastically, for example proliferating or dying according to local microenvironmental cues and probabilistic rules. In the present model, we focused on simplified, cancer cell–related processes driven by pulse intensity, to lay the foundations for a spatially resolved hybrid IBM. We thus restricted the present model to a limited set of cancer cell–related variables directly influenced by pulse intensity. Incorporating additional factors such as immune cells, ROS, or other complex biochemical feedback loops would have considerably increased the number of parameters and introduced uncertainties that could not be reliably constrained. Our aim at this stage was therefore to establish a robust, interpretable framework capturing the core tumour responses to electroporation, which can later be expanded in a modular way to include immune or oxidative components once validated datasets become available.

Performed \textit{in vitro} studies provided essential data on spheroid growth, death, and ATP and HMGB1 kinetics, supplying key elements required to build a predictive mathematical model, and its predictive power was validated on additional experimental experiments. In this study, the results of numerical simulations qualitatively reproduced the growth of tumour spheroids in control conditions (\textit{i.e.} prior the application of electroporation).  In particular, the model qualitatively reproduced the establishment of a proliferation gradient over time, characterized by a rising proportion of quiescent and necrotic cells in the spheroid centre. We have thereafter investigated the ability of our model to reproduce the effects of electroporation such as loss of viability and subsequent regrowth, as well as ATP and HMGB1 release. 

The quadratic regression model, employed to describe the relationship between electric field intensity and ATP levels in the supernatant showed a good fit to the experimental data.  By assuming that cell death was proportional to the amount of ATP released, that the duration before cell death takes place was proportional to the intensity of the electric field applied and that HMGB1 was secreted by dying cells, we validated the effectiveness of our model in qualitatively replicating the  dynamics  of  spheroid  death and HMGB1 release observed  in  the  experiments. In particular, our results reproduced the dynamics where high electric pulses induced rapid tumour cell death, causing cells exposed to a high electrical field intensity to secrete HMGB1 sooner than those exposed to lower intensities. Note that, to keep the model as simple as possible, we chose to include only mechanisms that were necessary to reproduce essential aspects of the experimental results. In future works, if spatio-temporal experimental measurements are available for ATP and HMGB1, we could calibrate the parameters related to their concentration dynamics, and then update the model in order to explicitly incorporate the dynamical modelling of these quantities using PDEs. Electroporation dynamics could also be represented through more detailed mathematical models, such as the one proposed in \cite{gallinato2019numerical, gallinato2020numerical}.

Finally, we have investigated the regrowth of the spheroids after electroporation. The results of numerical simulations capture the experimental phenomenon wherein lower electric field intensities have minimal impact on spheroid growth, intermediate intensities stimulate growth, and higher intensities lead to significant growth inhibition. For low intensities, this minimal effect can be attributed to the limited impact of electroporation on cells, which maintain dynamics similar to the control scenario. At intermediate field intensities, this behaviour is likely driven by increased death of proliferative and quiescent cells following electroporation. The loss of these cells frees up space and resources, allowing the surviving population to expand more rapidly and thereby accelerating tumour regrowth. Experimentally, this assumption was confirmed {\it a posteriori}, as all spheroids exposed to 1500 V/cm pulses showed uniform regrowth across the whole structure, even in the central region. Clinically, if tumour regrowth originates from residual viable cells in the core of the treated region, early recurrence might remain undetected in post-IRE imaging, as the proliferating cells would be embedded within the apparent ablation zone. This could contribute to the frequent observation of late local recurrences despite initially complete radiological responses. At high field intensities, the markedly reduced number of surviving proliferative cells after electroporation is insufficient to sustain normal growth, leading to spheroid growth inhibition. Overall, these results faithfully reproduce the experimentally observed sensitivity of spheroid dynamics to variations in electric field intensity and highlight the pivotal role of quiescent cells in driving tumour regrowth. On the one hand, quiescent cells that survived the treatment can re-enter the cell cycle and contribute to tumour regrowth. On the other hand, the model suggests that when quiescent cells die due to certain treatment, they free up space and resources, accelerating tumour regrowth by allowing surviving proliferative cells to expand more rapidly. Future experimental studies incorporating direct detection of proliferation markers (\textit{e.g.}, Ki-67, phospho-Histone H3, or BrdU incorporation) and flow cytometry-based cell cycle profiling would provide direct molecular confirmation of the regrowth dynamics observed here.

Taken together, the results of numerical simulations of our model indicate that all the assumptions we made were relevant and crucial for accurately reproducing the experimental results. These assumptions include factors such as the transition rates between proliferative, quiescent, and necrotic states, the effect of electric field intensity on ATP release, cell death, and its link with the dynamics of HMGB1 secretion. All these factors were validated through the close alignment of our simulation outcomes with observed experimental data. In particular, the assumption regarding the transitions between proliferative, quiescent, and necrotic states was essential. Without this assumption, it would have been difficult to reproduce the dynamics where moderate electric intensity stimulates regrowth without varying the parameters of tumour growth. This highlights the importance of our model’s comprehensive approach to accurately capture the complex biological responses to different electric field intensities.

While we managed to calibrate some parameters of the model from the literature (see Table \ref{ch4:table2}) and define them on the basis of precise biological assumptions, there are some parameters (e.g parameters related to the dynamics of the nutrient concentration, the switch rate at which proliferating and quiescent cells enter into necrotic state) whose values were chosen
	with a trial-and-error approach and to qualitatively reproduce essential aspects of the experimental
	results. In order to minimise the impact of this limitation on the conclusions of our study,  we first selected a baseline set of parameters that allowed to reproduce the growth
	of the spheroid area in control conditions. Subsequently, we carried out simulations by
	keeping all parameter fixed, with only the intensity of the electric pulse being varied. This variation influenced ATP release, which in turn affected tumour cell death, subsequently impacting HMGB1 release and cell regrowth. We finally compared the simulation results obtained through this process. We restrict our analysis to 2D tumour spheroids for several reasons. First, the experimental data used for comparison with the model were obtained from 2D images, making a 2D numerical framework more appropriate and straightforward for direct comparison. In addition, performing a single 3D simulation of a spheroid containing thousands of cells would require several hours to days of computational time. Such computational demands would make it impractical to conduct the extensive set of simulations presented in this study, particularly given that multiple simulations are needed for each parameter set. Furthermore, the 2D simulations considered here are sufficiently detailed to capture qualitative differences in cellular dynamics and molecular release patterns as parameters related to electroporation intensity are varied. Finally, establishing quantitative correspondences between tumour spheroid models implemented in 2D versus 3D, or across different computational frameworks, remains an active high-performance computing research topic and is beyond the scope of the present work. From a modelling point of view, although more tailored
to capture fine details of the dynamics of single cells, IBMs are not amenable to analytical studies, which may
support a more in-depth theoretical understanding of the application
problems under study. For this reason, in future work we plan
to derive a deterministic continuum model from a IBM by using mean filed methods similar to
those employed in \cite{almeida2022hybrid,chisholm2016evolutionary,painter2015navigating}. The use of continuum deterministic models would also reduce computational time and make it possible to perform more extensive sensitivity analyses of the parameters.

\section{Conclusions}\label{conclusion}

This interdisciplinary study combines experimental and computational approaches to investigate the temporal aspects of DAMP release induced by pulsed electric fields. Focusing on two specific DAMPs—ATP and HMGB1—it provides insights into how their dynamics may contribute to subsequent immune activation. The proposed hybrid individual-based model successfully replicates the effects of electroporation on tumour spheroids, including DAMP release, cell death, and cell regrowth. This study highlights the critical influence of electric pulse intensity, with higher intensities inhibiting re-growth and intermediate intensities stimulating it. Despite its relative simplicity, the proposed model offers a novel and cost-effective tool to explore therapeutic strategies involving electroporation, with the potential to inform the design of synergistic treatments combining pulsed electric field and immunotherapeutic strategies.


\section*{Acknowledgements} 

CP kindly acknowledges the Plan Cancer project MECI n◦21CM119-00 and the French national research agency (ANR) projects IMITATE (ANR-22-CE51-0043),  MIRE4VTACH (ANR-22-CE45-0014), and DEVIN (ANR-25-CE45-2591) which granted mathematical research. MG kindly thanks the FONROGA foundation (la fondation Roland Garrigou pour la culture et la santé) for the Master’s students grant to N.M. JKT kindly acknowledges the ANR funding, project JOULMECT (grant ANR-23-CE18-0029-01), founding biological experiments and the PhD grant to N.M. Authors also kindly thank the “Toulouse Reseau Imagerie” core IPBS facility (Genotoul, Toulouse, France).

 \section*{Author contributions}

EL: Methodology, Formal analysis, Software, Writing - review \& editing; NM: Biological experiments, Biological data analysis, Data curation,  Writing - review  \& editing; JKT: Investigation; Methodology; Supervision; Writing - review \& editing; Funding acquisition MPR: Funding acquisition; Project administration; MG: Investigation; Methodology; Supervision; Writing - review  \& editing; Funding acquisition; CP: Investigation; Methodology; Supervision; Funding acquisition; Project administration; Writing - review  \& editing.
\newpage

\bibliographystyle{unsrt}
\bibliography{biblio} 
\newpage
\appendix 
\clearpage
\appendix
\section*{Supporting information}
\addcontentsline{toc}{section}{Supplementary Material}


\renewcommand\thefigure{S\arabic{figure}}
\renewcommand\thetable{S\arabic{table}}
\renewcommand\theequation{S\arabic{equation}}
\setcounter{figure}{0}\setcounter{table}{0}\setcounter{equation}{0}
\section{Supplementary figures}
\begin{figure}[H]
	\centering
	\includegraphics[width=16cm]{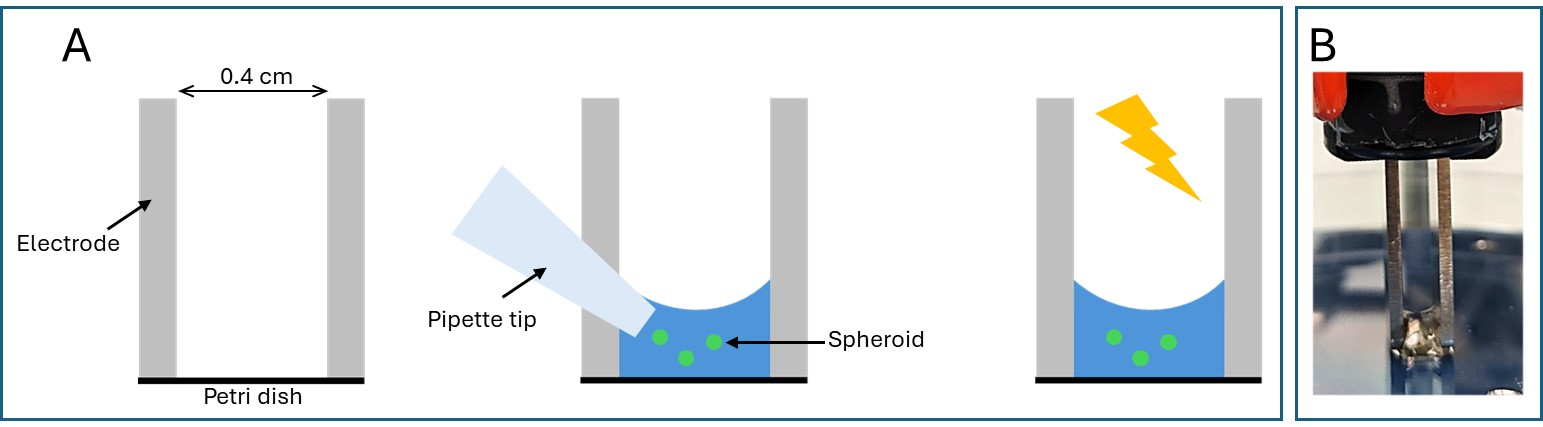}
	\caption{\textbf{Electroporation setting.}
		(A) Schematic representation of electrodes between which a spheroid-containing drop is placed and pulsed. (B) A photograph of the setting showing six pulsed drops and the (lifted) electrode.} 
	\label{supplementary2}
\end{figure}
\begin{figure}[H]
	\centering
	\includegraphics[width=16cm]{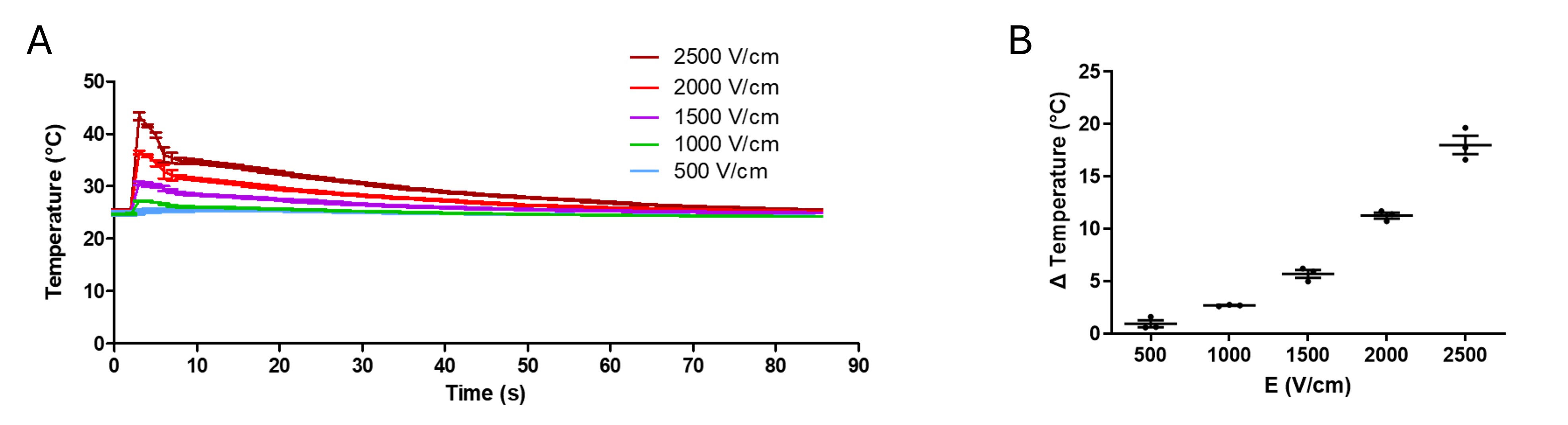}
	\caption{\textbf{Temperature increases related to the electroporation protocol.}
	(A) Temperature recordings obtained with a fiber-optic thermometer positioned within the droplet between the electrodes before pulse application and maintained in place during and for approximately 80 seconds after the pulsing protocols performed at different electric field intensities. (B) Corresponding maximal temperature increase ($\Delta$T) as a function of electric field intensity. Data represent the mean ± standard error of the mean (n = 3).}
	\label{supplementary3}
\end{figure}
\begin{figure}[H]
	\centering
	\includegraphics[width=7cm]{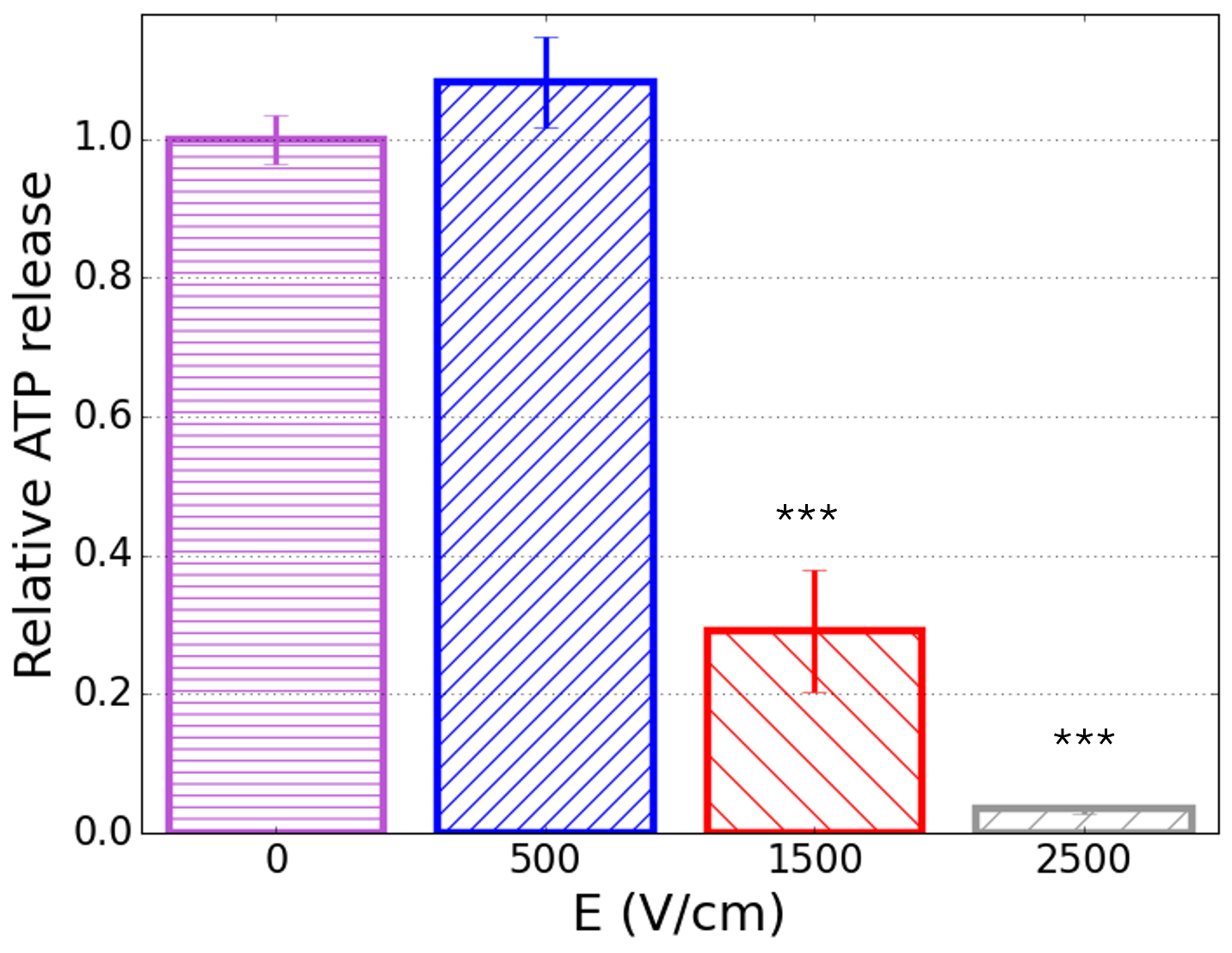}
	\caption{\textbf{Extractible ATP from IRE-treated spheroids.}
		 ATP level in supernatant measured 24 hours after pulse treatment of a pool of 3 spheroids (relative to ATP level at 0 V/cm). Statistical significance was assessed using an one-way ANOVA followed by a Dunnett’s multiple comparison test, in comparison to the 0 V/cm condition. *** $p < 0.001$. Data are shown as the means $\pm$ standard error mean from three independent experiments. }
	\label{supplementary}
\end{figure}

\section{Model calibration through parameter exploration} 
\label{calibration}
The model is calibrated to qualitatively reproduce the experimental results reported in section \ref{Material and method}. 
Due to the nature of the data (2D images of the spheroids, number of cells, amount of ATP and HMGB1 release) and to computational cost we developed a 2-dimensional version of our model. With our model, we  focused on qualitatively capturing the ATP and HMGB1 release and the regrowth of tumour spheroids while varying only the intensity of the electric field applied to the spheroids. 

Some parameters of the model (see Table \ref{ch4:table2}) are estimated from the literature
and defined on the basis of precise biological assumptions. Other model parameters that could not be based on a literature source, such as the switching rate from proliferative/quiescent cells to necrotic cells, are identified using a trial-and-error approach to qualitatively reproduce the growth of the spheroids  in normal conditions observed in the experiments. A sensitivity analysis of our model to some parameters (\textit{e.g.} $n_{max}$, $\beta$, $\gamma_{min}$, $C_{hyp}$) is presented in Appendix \ref{sensitivity analysis}.
\begin{table}[hp!]
	\small
\caption{Parameter values used in numerical simulations.}
\centering
\begin{tabular}{p{1.8cm} p{4.4cm} p{2.4cm}p{4.3cm} p{1cm}}
 \hline
  Phenotype  & Description & Symbol & Value & Ref. \\
\hline
\textbf{Domain} 
& Domain  & $\overline{\Omega}$ & $[-1, 1]^2$ [\textrm{mm$^2$}]&Calibrated \\ 
& Space-step & $\chi$ & 0.033 [\textrm{mm}] & Calibrated\\ 
& Time-step & $\tau$ & 0.005 [\textrm{days}] &Calibrated\\ 
\textbf{Proliferative} & Cell density at position $\mathbf{x_i}$ and time $t_k$  & $p_\mathbf{i}^k=p(\mathbf{x_i},t_k)$ & values $\geq 0$ [\textrm{cells/mm$^2$}] &\\ 
& Proliferation rate   & $\alpha_p$ & 1 [\textrm{1/day}]
& \cite{leschiera2023individual}
\\ 
\textbf{Quiescent} & Cell density at position $\mathbf{x_i}$ and time $t_k$  & $q_\mathbf{i}^k=q(\mathbf{x_i},t_k)$ & values $\geq 0$ [\textrm{cells/mm$^2$}] &\\ 
\textbf{Necrotic} & Cell density at position $\mathbf{x_i}$ and time $t_k$  & $r_\mathbf{i}^k=r(\mathbf{x_i},t_k)$ & values $\geq 0$ [\textrm{cells/mm$^2$}] &\\
& Switching rate from proliferative/quiescent to necrotic  & $\gamma(C_\mathbf{i}^k)$ & see equation \eqref{switching}&\cite{michel2018mathematical}\\
&   & $\gamma_{\min}$ &$0$ & Calibrated\\
&   & $\gamma_{\max}$ &$0.6$&Calibrated \\
&   & $K$ &$5$ &Calibrated\\
&   & $C_{\text{hyp}}$ &$0.7$ &Calibrated\\
\textbf{Dead} & Cell density at position $\mathbf{x_i}$ and time $t_k$  & $m_\mathbf{i}^k=m(\mathbf{x_i},t_k)$ & values $\geq 0$ [\textrm{cells/mm$^2$}] &\\ 
\textbf{Tumour} & Cell density at position $\mathbf{x_i}$ and time $t_k$  & $n_\mathbf{i}^k=n(\mathbf{x_i},t_k)$ & values $\geq 0$ [\textrm{cells/mm$^2$}]&\\
 &Area at time $t_k$  & $AS^k=AS(t_k)$ & values $\geq 0$ [mm$^2$] &\\
  & Cell density above which cell proliferation is impaired &$n_{\max}$ &  $4.875 \times 10^5$ [\textrm{cells/mm$^2$}] & Calibrated \\
  & Cell death rate due to electroporation & $p$& $p=c_1Z$ \\
  &   & $c_1$ &$0.0255$ & Calibrated\\
  &   & $Z$ &$\mathcal{U}_{[S-10\%S,S+10\%S]}$ \\
\textbf{Nutrient} & Concentration at position $\mathbf{x_i}$ and time $t_k$  & $C_\mathbf{i}^k=C(\mathbf{x_i},t_k)$ & values $\geq 0$ [\textrm{mol/mm$^2$}] &\\
 & Initial concentration  & $C_\mathit{init}$ & 5000  [\textrm{mol/mm$^2$}] &
 \\ 
 & Diffusion coefficient  & $\beta_C$ & $1.5\times 10^{-2}$  [\textrm{mm$^2$/days}] & Calibrated 
 \\
  & Consumption rate by proliferative cells  & $\alpha_C$ & $2\times 10^{-6}$ [\textrm{mol/(cells days)}]  & Calibrated
  \\
   & Consumption rate by quiescent cells  & $\eta_C$ & $c_2\alpha_C$  [\textrm{mol/(cells days)}] & Calibrated\\ 
   &  & $c_2$&$\frac{2}{3}$ & Calibrated
   \\
\textbf{ATP} &Amount in the supernatant 10 min. after electroporation   & $S( \bold{E})$ & $\beta_0 + \beta_1 \bold{E} + \beta_2 \bold{E}^2$ [\textrm{mol}]\\
     &  & $\beta_0$ & 1524  & Estimated\\
      &  & $\beta_1$ & 486 & Estimated \\
       &  & $\beta_2$ & 3.9 & Estimated\\
        \textbf{HMGB1} &Amount in the supernatant at time $t_k$  & $P^k=P(t_k)$  & values $\geq 0$ (\textrm{mol}) \\
        &Release rate by dying cells  & $\alpha_P$ & 0.2 [\textrm{mol/(cells days)}] &Calibrated\\
\hline
\end{tabular}
\label{ch4:table2}
\end{table}
\section{Sensitivity of the model}
\label{sensitivity analysis}
We managed to determine some of the parameters of the model (see Tables \ref{ch4:table2}) from the literature and also by making reasonable biological assumptions. Nevertheless,
we still rely upon the value of some free parameters (\textit{e.g.} $n_{max}$, $\alpha_C$, $\beta_C$, $\eta_C$, $\gamma_{min}$, $\gamma_{max}$, $C_{hyp}$), which could not properly be derived
from the literature. We therefore studied the impact of some of those free parameters, namely $n_{max}$, $\alpha_C$, $\gamma_{min}$ and $C_{hyp}$
on the simulation outcomes. Parameter $n_{max}$  characterizes the cell density above which cell proliferation is impaired, $\alpha_C$ is the nutrient consumption  rate  by  proliferative cells (which also affects the consumption  rate  by  quiescent cells, \textit{cf.} equation \eqref{nutrent}) and $\gamma_{min}$ and $C_{hyp}$ are two parameters of equation \eqref{switching} describing the switch rate at which proliferating and quiescent cells enter the necrotic state. These parameters act directly on the transition between proliferative, quiescent and necrotic state and can then be considered as essential for the definition of our control scenario. 

\begin{figure}[H]
	\centering
	\includegraphics[width=15cm]{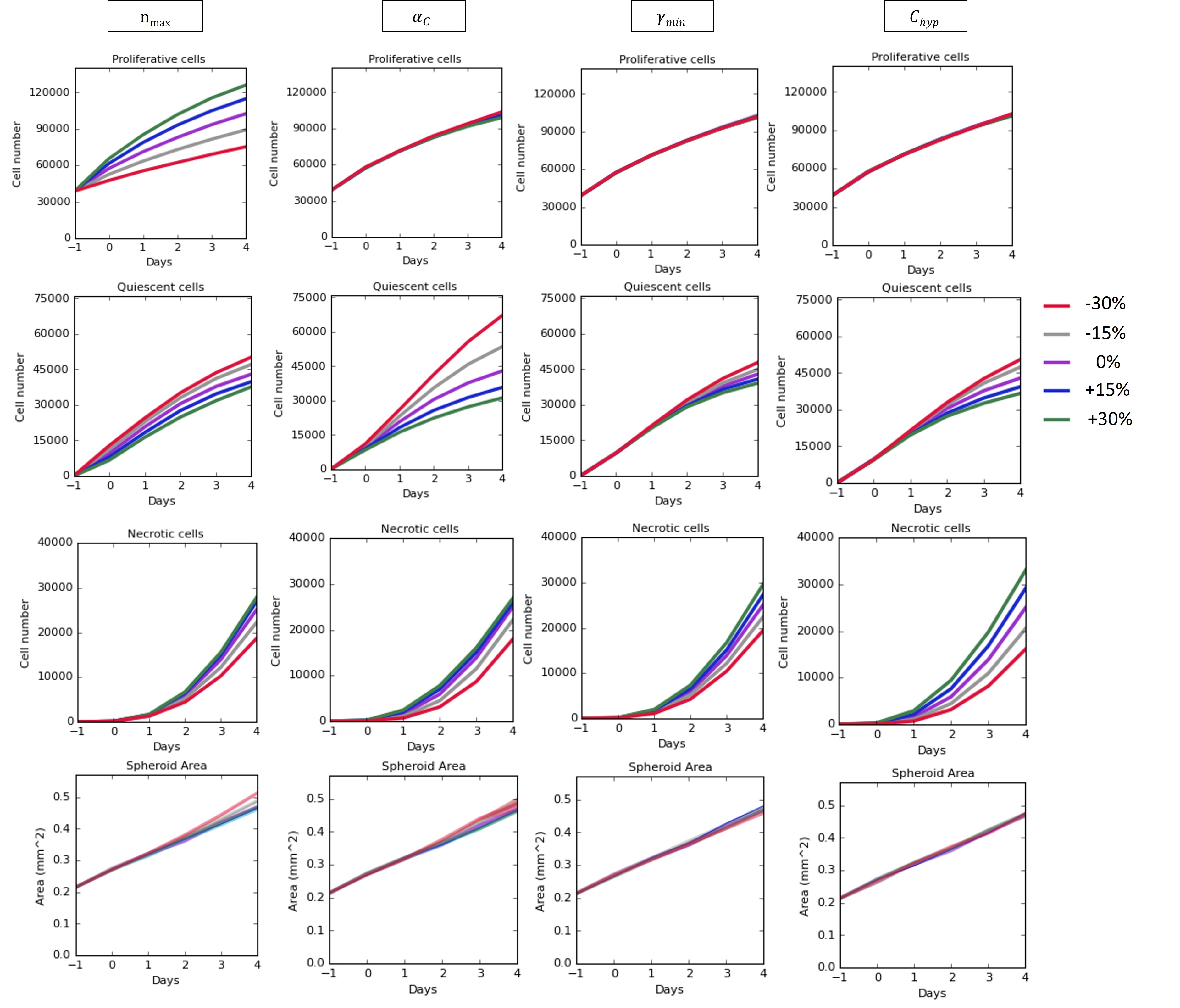}
	\caption{\textbf{Sensitivity of the model to parameters $n_{max}$, $\alpha_C$, $\gamma_{min}$, $C_{hyp}$.}
		Time evolution of the proliferative, quiescent and necrotic cell number, as well as the time evolution of the spheroid area,  for variations of $-30\%, -15\%, +15\%, +30\%$ from the baseline values of parameters $n_{max}$, $\alpha_C$, $\gamma_{min}$, $C_{hyp}$. These results correspond to the average over 3 simulations and the shaded area indicates +/- standard deviation.} 
		\label{sensitivity}
\end{figure}

Figure~\ref{sensitivity} shows the number of proliferative, quiescent and necrotic cells, as well as the spheroid area for different values of  these four parameters of interest. In particular, it shows that parameter $n_{max}$ affects both proliferative, quiescent and necrotic cell dynamics. In fact, as this parameter increases, the number of proliferative and necrotic cells increases, while the number of quiescent cells decreases. This is due to the fact that higher values of $n_{max}$ provide more space for cell proliferation, which reduces the transition of proliferative cells to a quiescent state and increases the transition of proliferative cells to a necrotic state. Conversely, when this parameter is decreased, the opposite effect is observed.

Variations in parameter $\alpha_C$ primarily affect the dynamics of quiescent and necrotic cells. Specifically, Figure~\ref{sensitivity} shows that the increase of $\alpha_C$ leads to a decrease of quiescent cells and an increase of necrotic cells. This is due to the fact that higher values of $\alpha_C$ result in lower nutrient concentration, increasing the probabilities of proliferative and quiescent cells transitioning to necrotic states. In particular, Figure~\ref{sensitivity} shows that higher values of $\alpha_C$ particularly increase the transition from quiescent to necrotic states rather than from proliferative to quiescent states. This is likely due to the fact that proliferative cells have a greater likelihood of transitioning to quiescent states before becoming necrotic.

Finally, Figure~\ref{sensitivity} shows that variations in parameters $\gamma_{min}$ and $C_{hyp}$ impact tumour cell dynamics in a similar way as parameter $\alpha_C$. As shown by Figure~\ref{sensitivity}, the increase of these two parameters results in a decrease of quiescent cells and an increase of necrotic cells.  This is due to the fact that higher values of these parameters accelerate the switch rate at which proliferating and quiescent cells enter the necrotic state. As for parameter $\alpha_C$, Figure~\ref{sensitivity} shows that higher values of $\gamma_{min}$ and $C_{hyp}$ particularly increase the transition from quiescent to necrotic states rather than from proliferative to quiescent states.

It is important to note that the spheroid area is almost not affected by modifications to these four parameters, as an increase in one of the three tumour states (proliferative, quiescent, or necrotic) is compensated by a decrease in another.

\end{document}